\documentclass[11pt, oneside]{article}   
\usepackage[utf8]{inputenc}
\usepackage{amsmath,amssymb,amsthm,algorithmicx,algorithm,setspace,url}
\usepackage{graphicx,bbm}
\usepackage[top=0.6in,margin=1in]{geometry}
\usepackage[authoryear, sort]{natbib}
\usepackage{authblk,booktabs,epstopdf,color,lscape,float}
\usepackage[hidelinks]{hyperref}
\usepackage{multirow}
\usepackage[space]{grffile}
\date{}
\def\Gauss{{\mathrm{N}}}

\def\mr{\mathrm}

\def\HS{{\textbf{HS}}}
\def\T{\mathrm{\scriptscriptstyle{T}}}

\newtheorem{theorem}{Theorem}[section]
\newtheorem{lemma}[theorem]{Lemma}
\newtheorem{assumption}[theorem]{Assumption}
\newtheorem{proposition}[theorem]{Proposition}
\newtheorem{corollary}[theorem]{Corollary}

\newcommand{\norm}[1]{\left\lVert #1 \right\rVert}
\DeclareMathOperator*{\argmin}{arg\,min}
\mathchardef\mhyphen="2D
\begin{document}
\thispagestyle{empty}
\baselineskip=28pt
\begin{center}
{\LARGE{\bf  Bayesian sparse multiple regression for simultaneous rank reduction and variable selection}}

\end{center}
\baselineskip=12pt

\vskip 10mm
\begin{center}
Antik Chakraborty\\
Department of Statistics, Texas A\&M University, College Station\\
3143 TAMU, TX 77843-3143, USA\\
antik@stat.tamu.edu\\
\hskip 5mm\\
Anirban Bhattacharya\\
Department of Statistics, Texas A\&M University, College Station\\
3143 TAMU, TX 77843-3143, USA\\
anirbanb@stat.tamu.edu\\
\hskip 5mm\\
Bani K. Mallick\\
Department of Statistics, Texas A\&M University, College Station\\
3143 TAMU, TX 77843-3143, USA\\
bmallick@stat.tamu.edu\\

\end{center}

\baselineskip=22pt

\vskip 10mm
\begin{center}
{\Large{\bf Abstract}}
\end{center}
\baselineskip=14pt
We develop a Bayesian methodology aimed at simultaneously estimating low-rank and row-sparse matrices in a high-dimensional multiple-response linear regression model. We consider a carefully devised shrinkage prior on the matrix of regression coefficients which obviates the need to specify a prior on the rank, and shrinks the regression matrix towards low-rank and row-sparse structures. We provide theoretical support to the proposed methodology by proving minimax optimality of the posterior mean under the prediction risk in ultra-high dimensional settings where the number of predictors can grow sub-exponentially relative to the sample size. A one-step post-processing scheme induced by group lasso penalties on the rows of the estimated coefficient matrix is proposed for variable selection, with default choices of tuning parameters. We additionally provide an estimate of the rank using a novel optimization function achieving dimension reduction in the covariate space. We exhibit the performance of the proposed methodology in an extensive simulation study and a real data example.
\vspace{1.5in}

\baselineskip=12pt
\par\vfill\noindent

\par\medskip\noindent
\underline{\bf Short title}: Bayesian sparse multi-task learner

\clearpage\pagebreak\newpage
\pagenumbering{arabic}
\newlength{\gnat}
\setlength{\gnat}{22pt}
\baselineskip=\gnat

\section{Introduction}

Studying the relationship between multiple response variables and a set of predictors has broad applications ranging from bioinformatics, econometrics, time series analysis to growth curve models. The least squares solution in a linear multiple response regression problem is equivalent to performing separate least squares on each of the responses \citep{anderson1984multivariate} and ignores any potential dependence among the responses. 
In the context of multiple response regression, a popular technique to achieve parsimony and interpretability is to consider a reduced-rank decomposition of the coefficient matrix, commonly known as reduced rank regression \citep{anderson1951estimating,izenman1975reduced,velu2013multivariate}. 
Although many results exist about the asymptotic properties of reduced rank estimators \citep{anderson2002specification}, formal statistical determination of the rank remains difficult even with fixed number of covariates and large sample size due mainly to the discrete nature of the parameter. 
The problem becomes substantially harder when a large number of covariates are present, and has motivated a series of recent work on penalized estimation of low rank matrices, where either the singular values of the coefficient matrix \citep{yuan2007dimension,chen2013reduced}, or the rank itself \citep{bunea2011optimal} is penalized. 
Theoretical evaluations of these estimators focusing on adaptation to the oracle convergence rate when the true coefficient matrix is of low rank has been conducted in \cite{bunea2011optimal}. It has also been noted \citep{bunea2012joint} that the convergence rate can be improved when the true coefficient matrix has zero rows and variable selection is incorporated within the estimation procedure. Methods that simultaneously handle rank reduction and variable selection include \cite{yuan2007dimension,bunea2012joint,chen2012sparse}. To best of our knowledge, uncertainty characterization for the parameter estimates from these procedures is currently not available. 

The first fully systematic Bayesian treatment of reduced rank regression was carried out in \cite{geweke1996bayesian}, where  conditioned on the rank, independent Gaussian priors were placed on the elements of the coefficient matrix. While formal Bayesian model selection can be performed to determine the rank \citep{geweke1996bayesian}, calculation of marginal likelihoods for various candidate ranks gets computationally burdensome with increasing dimensions.
The problem of choosing the rank is not unique to reduced rank regression and is ubiquitous in situations involving low rank decompositions, with factor models being a prominent example. \cite{lopes2004bayesian} placed a prior on the number of factors and proposed a computationally intensive reversible jump algorithm \citep{green1995reversible} for model fitting. 
As an alternative, \cite{bhattacharya2011sparse} proposed to increasingly shrink the factors starting with a conservative upper bound and adaptively collapsing redundant columns inside their MCMC algorithm.
Recent advancements in Bayesian matrix factorization have taken a similar approach; see for example, \cite{lim2007variational,salakhutdinov2008bayesian,babacan2011variational,alquier2013bayesian}. 

From a Bayesian point of view, a natural way to select variables in a single-response regression framework is to use point mass mixture priors \citep{george1993variable,scott2010bayes} which allow a subset of the regression coefficients to be exactly zero. These priors were also adapted to multiple response regression by several authors \citep{brown1998multivariate,lucas2006sparse,wang2010sparse,bhadra2013joint}.
Posterior inference with such priors involves a stochastic search over an exponentially growing model space and is computationally expensive even in moderate dimensions. To alleviate the computational burden, a number of continuous shrinkage priors have been proposed in the literature which mimic the operating characteristics of the discrete mixture priors. Such priors can be expressed as Gaussian scale mixtures \citep{polson2010shrink}, leading to block updates of model parameters; see \cite{bhattacharya2016fast} for a review of such priors and efficient implementations in high-dimensional settings. To perform variable selection with these continuous priors, several methods for post-processing the posterior distribution have been proposed \citep{bondell2012consistent,kundu2013bayes,hahn2015decoupling}.


In this article we simultaneously address the problems of dimension reduction and variable selection in high-dimensional reduced rank models from a Bayesian perspective. We develop a novel shrinkage prior on the coefficient matrix which encourages shrinkage towards low-rank and row-sparse matrices. The shrinkage prior is induced from appropriate shrinkage priors on the components of a full-rank decomposition of the coefficient matrix, and hence bypasses the need to specify a prior on the rank. We provide theoretical understanding into the operating characteristics of the proposed prior in terms of a novel prior concentration result around rank-reduced and row-sparse matrices. The prior concentration result is utilized to prove minimax concentration rates of the posterior under the fractional posterior framework of \cite{bhattacharya2016bayesian} in a ultrahigh-dimensional setting where the number of predictor variables can grow sub-exponentially in the sample size. 

The continuous nature of the prior enables efficient block updates of parameters inside a Gibbs sampler. In particular, we adapt an algorithm for sampling structured multivariate Gaussians from \cite{bhattacharya2016fast} to efficiently sample a high-dimensional matrix in a block leading to a low per-iteration MCMC computational cost. We propose two independent post-processing schemes to achieve row sparsity and rank reduction with encouraging performance. A key feature of our post-processing schemes is to exploit the posterior summaries to offer careful default choices of tuning parameters, resulting in a procedure which is completely free of tuning parameters. The resulting row-sparse and rank-reduced coefficient estimate is called a Bayesian sparse multi-task learner (BSML). We illustrate the superiority of BSML over its competitors through a detailed simulation study and the methodology is applied to a Yeast cell cycle data set. Code for implementation is available at \url{www.stat.tamu.edu/~antik}. 


\section{Bayesian sparse multitask learner}
\subsection{Model and Prior Specification}\label{sec:mod}
Suppose, for each observational unit $i = 1, \ldots, n$, we have a multivariate response $y_i \in \Re^q$ on $q$ variables of interest, along with information on $p$ possible predictors $x_i \in \Re^p$, a subset of which are assumed to be important in predicting the $q$ responses. Let $X \in \Re^{n \times p}$ denote the design matrix whose $i$th row is $x_i^{\T}$, and $Y \in \Re^{n\times q}$ the matrix of responses with the $i$th row as $y_i^{\T}$. The multivariate linear regression model is,
\begin{align}\label{eq:lin_reg}
Y=XC+E, \hspace{0.2in}E=(e_1^\T,\ldots,e_n^\T)^\T,
\end{align}
where we follow standard practice to center the response and exclude the intercept term. The rows of the error matrix are independent, with $e_i \sim \Gauss(0,\Sigma)$. Our main motivation is the high-dimensional case where $p \ge \max\{n, q\}$, although the method trivially applies to $p < n$ settings as well. We shall also assume the dimension of the response $q$ to be modest relative to the sample size. 

The basic assumption in reduced rank regression is that $\text{rank}(C)=r\leq \mathrm{min}(p,q)$, whence $C$ admits a decomposition $C = B_*A_*^\T$ with $B_* \in \Re^{p \times r}$ and $A_* \in \Re^{q \times r}$. While it is possible to treat $r$ as a parameter and assign it a prior distribution inside a hierarchical formulation, posterior inference on $r$ requires calculation of intractable marginal likelihoods or resorting to complicated reversible jump Markov chain Monte Carlo algorithms. To avoid specifying a prior on $r$, we work within a parameter-expanded framework \citep{liu1999parameter} to consider a potentially full-rank decomposition $C = B A^{\T}$ with $B\in \Re^{p \times q}$ and $A \in \Re^{q \times q}$, and assign shrinkage priors to $A$ and $B$ to shrink out the redundant columns when $C$ is indeed low rank. This formulation embeds all reduced-rank models inside the full model; if a conservative upper bound $q^\ast \le q$ on the rank is known, the method can be modified accordingly. The role of the priors on $B$ and $A$ is important to encourage appropriate shrinkage towards reduced-rank models, which is discussed below. 

We consider independent standard normal priors on the entries of $A$. As an alternative, a uniform prior on the Stiefel manifold \citep{hoff2007gibbs} of orthogonal matrices can be used. However, our numerical results suggested significant gains in computation time using the Gaussian prior over the uniform prior with no discernible difference in statistical performance. The Gaussian prior allows an efficient block update of $\mbox{vec}(A)$, whereas the algorithm of \cite{hoff2007gibbs} involves conditional Gibbs update of each column of $A$. Our theoretical results also suggest that the shrinkage provided by the Gaussian prior is optimal when $q$ is modest relative to $n$, the regime we operate in. We shall henceforth denote $\Pi_A$ to denote the prior on $A$, i.e., $a_{hk} \sim \mbox{N}(0, 1)$ independently for $h, k = 1, \ldots, q$. 

Recalling that the matrix $B$ has dimension $p \times q$, with $p$ potentially larger than $n$, stronger shrinkage is warranted on the columns of $B$. We use independent horseshoe priors \citep{carvalho2010horseshoe} on the columns of $B$, which can be represented hierarchically as 
\begin{align}\label{eq:HS}
b_{jh} \mid \lambda_{jh},\tau_h \sim \Gauss(0,\lambda_{jh}^2 \tau_h^2), \quad \lambda_{jh} \sim \mbox{Ca}_+(0, 1), \quad \tau_h \sim \mbox{Ca}_+(0, 1),
\end{align}
independently for $j = 1, \ldots, p$ and $h = 1, \ldots, q$, where $\mbox{Ca}_+(0, 1)$ denotes the truncated standard half-Cauchy distribution with density proportional to $(1+t^2)^{-1} \mathbbm{1}_{(0, \infty)}(t)$. We shall denote the prior on the matrix $B$ induced by the hierarchy in \eqref{eq:HS} by $\Pi_B$.

We shall primarily restrict attention to settings where $\Sigma$ is diagonal, $\Sigma = \mbox{diag}(\sigma_1^2, \ldots, \sigma_q^2)$, noting that extensions to non-diagonal $\Sigma$ can be incorporated in a straightforward fashion. For example, for moderate $q$, a conjugate inverse-Wishart prior can be used as a default. Furthermore, if $\Sigma$ has a factor model or Gaussian Markov random field structure, they can also be incorporated using standard techniques \citep{bhattacharya2011sparse,rue2001fast}. The cost-per-iteration of the Gibbs sampler retains the same complexity as in the diagonal $\Sigma$ case; see \S \ref{section:post_comp} for more details. In the diagonal case, we assign independent improper priors $\pi(\sigma_h^2) \propto \sigma_h^{-2}, \,h = 1, \ldots, q$ on the diagonal elements, and call the resulting prior $\Pi_{\Sigma}$. 

The model augmented with the above priors now takes the shape 
\begin{align}
&Y = XBA^\T+E, \quad e_i \sim \Gauss(0,\Sigma), \label{eq:c_factor} \\
& B \sim \Pi_B, \quad A \sim \Pi_A, \quad \Sigma \sim \Pi_{\Sigma}. \label{eq:prior}
\end{align}
We shall refer to the induced prior on $C = B A^{\T}$ by $\Pi_C$, and let 
$$
p^{(n)}(Y \mid C, \Sigma; X) \propto  |\Sigma|^{-n/2} \, e^{- \mbox{tr}\{ (Y - X C)\Sigma^{-1} (Y - X C)^{\T} \}/2 }
$$
denote the likelihood for $(C, \Sigma)$. 

\section{Posterior Computation}\label{section:post_comp}
Exploiting the conditional conjugacy of the proposed prior, we develop a straightforward and efficient Gibbs sampler to update the model parameters in \eqref{eq:c_factor} from their full conditional distributions. We use vectorization to update parameters in blocks. Specifically, in what follows, we will make multiple usage of the following identity. For matrices $\Phi_1, \Phi_2, \Phi_3$ with appropriate dimensions, and $\text{vec}(A)$ denoting column-wise vectorization, we have, 
\begin{equation}\label{eq:vectorization}
\text{vec}(\Phi_1\Phi_2\Phi_3)=(\Phi_3^\T \otimes \Phi_1)\text{vec}(\Phi_2)=(\Phi_3^\T \Phi_2^\T \otimes I_k)\text{vec}(\Phi_1),
\end{equation}
where the matrix $\Phi_1$ has $k$ rows and $\otimes$ denotes the Kronecker product. 

Letting $\theta\mid -$ denote the full conditional distribution of a parameter $\theta$ given other parameters and the data, the Gibbs sampler cycles through the following steps, sampling parameters from their full conditional distributions: 

\noindent \textbf{Step 1. } To sample $B \mid -$, use \eqref{eq:vectorization} to vectorize $Y=XBA^\T+E$ to obtain,
\begin{equation}\label{eq:vec_lin_reg1}
y =(X\otimes A) \beta + e, 
\end{equation} 
where $\beta = \text{ vec}(B^\T) \in \Re^{pq \times 1}$, $y = \text{vec}(Y^\T) \in \Re^{n q \times 1}$, and $e=\text{vec}(E^\T) \sim \Gauss_{nq}(0,\widetilde{\Sigma})$ with $\widetilde{\Sigma}=\text{diag}(\Sigma,\ldots,\Sigma)$. Multiplying both sides of \eqref{eq:vec_lin_reg1} by $\widetilde{\Sigma}^{-1/2}$ yields $\widetilde{y}=\widetilde{X}\beta+\widetilde{e}$ where $\widetilde{y}=\widetilde{\Sigma}^{-1/2} y$, $\widetilde{X}=\widetilde{\Sigma}^{-1/2}(X \otimes A)$ and $\widetilde{e}=\widetilde{\Sigma}^{-1/2}e \sim \Gauss_{nq}(0,\mathrm{I}_{nq})$. Thus, the full conditional distribution $\beta \mid - \sim \Gauss_{pq}(\Omega_B^{-1}\widetilde{X}^\T\widetilde{y},\Omega_B^{-1})$, where $\Omega_B=(\widetilde{X}^\T \widetilde{X}+\Lambda^{-1})$ with $\Lambda=\text{diag}(\lambda_{11}^2\tau_1^2,\ldots,\lambda_{1q}^2\tau_q^2,\ldots,\lambda_{p1}^2\tau_1^2,\ldots,\lambda_{pq}^2\tau_q^2)$. 

Naively sampling from the full conditional of $\beta$ has complexity $O(p^3q^3)$ which becomes highly expensive for moderate values of $p$ and $q$. \cite{bhattacharya2016fast} recently developed an algorithm to sample from a class of structured multivariate normal distributions whose complexity scales linearly in the ambient dimension. 
We adapt the algorithm in \cite{bhattacharya2016fast} as follows: 

\,

 \noindent (i) Sample $u \sim \Gauss(0,\Lambda)$ and $\delta \sim \Gauss(0,\mr I_{nq})$ independently. \\
   \noindent (ii) Set $v=\widetilde{X} u+\delta$.\\
   \noindent (iii) Solve $(\widetilde{X} \Lambda\widetilde{X}^{\T}+\mr I_{nq})w = (\tilde{y} - v)$ to obtain $w$.\\
   \noindent (iv) Set $\beta=u+\Lambda \widetilde{X}^{\T}w$.

It follows from \cite{bhattacharya2016fast} that $\beta$ obtained from steps (i) - (iv) above produce a sample from the desired full conditional distribution. One only requires matrix multiplications and linear system solvers to implement the above algorithm, and no matrix decomposition is required. It follows from standard results \citep{golub2012matrix} that the above steps have a combined complexity of $O(q^3 \max\{n^2, p\})$, a substantial improvement over $O(p^3 q^3)$ when $p \gg \max\{n, q\}$. 

\,
\noindent \textbf{Step 2.} To sample $A \mid -$, once again vectorize $Y=XBA^\T+E$, but this time use the equality of the first and the third terms in \eqref{eq:vectorization} to obtain,
\begin{equation}\label{eq:vec_lin_reg2}
y =(XB \otimes \mathrm{I}_q) a + e,
\end{equation}
where $e$ and $y$ are the same as in step 1, and $a = \text{vec}(A) \in \Re^{q^2 \times 1}$. The full conditional posterior distribution $a \mid - \sim \Gauss(\Omega_A^{-1}X_*\widetilde{y},\Omega_A^{-1})$,  where $\Omega_A = (X_*^\T X_*+\mathrm{I}_{q^2})$, $X_*=\widetilde{\Sigma}^{-1/2}(XB \otimes \mathrm{I}_{q^2})$ and $\widetilde{y} = \widetilde{\Sigma}^{-1/2} y$. To sample from the full conditional of $a$, we use the algorithm from \S 3.1.2 of  \cite{rue2001fast}. Compute the Cholesky decomposition $(X_*^{\T} X_* + I_{q^2}) = LL^\T$. Solve the system of equations: $L v = X_*^{\T}\tilde{y}$, $L^{\T} m = v$, and $L^{\T} w = z$, where $z \sim \Gauss(0,\mr{I_{q^2}})$. Finally obtain a sample as $a=m+w$.
\,

\noindent \textbf{Step 3.} To sample $\sigma_h^2 \mid -$, observe that $\sigma_h^2 \mid - \sim \text{inverse-Gamma}(n/2,S_h/2)$ independently across $h$, where $S_h=\{ Y_h-(XBA^\T)_h \}^\T \{ Y_h-(XBA^\T)_h \}$, with $\Phi_h$ denoting the $h$th column of a matrix $\Phi$. In the case of an unknown $\Sigma$ and an inverse-Wishart$(q,\mathrm{I}_q)$ prior on $\Sigma$, the posterior update of $\Sigma$ can be easily modified due to conjugacy; we sample $\Sigma\mid -$ from inverse-Wishart$\{n+q,(Y-XC)^\T (Y-XC)+\mathrm{I}_q\}$. 

\, 

\noindent \textbf{Step 4.} The global and local scale parameters $\lambda_{jh}$'s and $\tau_h$'s have independent conditional posteriors across $j$ and $h$, which can be sampled via a slice sampling scheme provided in the online supplement to \cite{polson2014bayesian}. We illustrate the sampling technique for a generic local shrinkage parameter $\lambda_{jh}$; a similar scheme works for $\tau_h$. 
Setting $\eta_{jh} = \lambda_{jh}^{-2}$, the slice sampler proceeds by sampling $u_{jh} \mid \eta_{jh} \sim \text{Unif}(0,1/(1+\eta_{jh}))$ and then sampling $\eta_{jh} \mid u_{jh} \sim \text{Exp}(2\tau_h^2/b_{jh}^2)\mathrm{I}\{ \eta_{jh}<(1-u_{jh})/u_{jh} \}$, a truncated exponential distribution. 

The Gibbs sampler above when modified to accommodate non-diagonal $\Sigma$ as mentioned in step 3 retains the overall complexity. Steps 1-2 do not assume any structure for $\Sigma$. The matrix $\Sigma^{-1/2}$ can be computed in $O(q^3)$ steps using standard algorithms, which does not increase the overall complexity of steps 1 and 2 since since $q<n\ll p$ by assumption. Modifications to situations where $\Sigma$ has a graphical/factor model structure are also straightforward. 

Point estimates of $C$, such as the posterior mean, or element-wise posterior median, are readily obtained from the Gibbs sampler along with a natural uncertainty quantification, which can be used for point and interval predictions. However, the continuous nature of our prior implies that such point estimates will be non-sparse and full rank with probability one, and hence not directly amenable for variable selection and rank estimation. Motivated by our concentration result in Theorem \ref{th:theorem2} that the posterior mean $X \overline{C}$ increasingly concentrates around $X C_0$, we propose two simple post-processing schemes for variable selection and rank estimation below. The procedures are completely automated and do not involve any input of tuning parameters from the user's end. 

\subsection{Post processing for variable selection}
We first focus on variable selection. We define a row-sparse estimate $\widehat{C}_R$ for $C$ as the solution to the optimization problem  
\begin{equation}\label{eq:utility12}
\widehat{C}_R=\argmin_{\Gamma \in \Re^{p \times q}} \bigg\{ \| X\overline{C}-X\Gamma \|_F^2 + \sum_{j=1}^p \mu_j \| \Gamma^{(j)} \|_2 \bigg\},
\end{equation} 
where $\Phi^{(j)}$ represents the $j^{th}$ row of a matrix $\Phi$, and the $\mu_j$s are predictor specific regularization parameters. The objective function aims to find a row-sparse solution close to the posterior mean in terms of the prediction loss, with the sparsity driven by the group lasso penalty \citep{yuan2006model}. For a derivation of the objective function in \eqref{eq:utility12} from a utility function perspective as in \cite{hahn2015decoupling}, refer to the supplementary document. 

%
%
%


To solve \eqref{eq:utility12}, we set the sub-gradient of \eqref{eq:utility12} with respect to $\Gamma^{(j)}$ to zero and replace $\|\Gamma^{(j)} \|$ by a data dependent quantity to obtain the soft thresholding estimate,
\begin{equation}\label{eq:subgrad2}
\widehat{C}_R^{(j)}=\dfrac{1}{X_j^\T X_j}\left(1-\dfrac{\mu_j}{2\| X_j^\T R_j \|}\right )_+ X_j^\T R_j, 
\end{equation}
where for $x\in \Re, x_+=\max(x,0)$, and $R_j$ is the residual matrix obtained after regressing $X\overline{C}$ on $X$ leaving out the $j^{th}$ predictor, $R_j= X \overline{C} - \sum_{k\neq j}X_k \widehat{C}_R^{(k)}$. See the supplementary document for the derivation of \eqref{eq:subgrad2}. For practical implementation, we use $\overline{C}$ as our initial estimate and make a single pass through each variable to update the initial estimate according to \eqref{eq:subgrad2}. With this initial choice, $R_j=X_j\overline{C}^{(j)}$ and $\| X_j^\T R_j\|=\|X_j\|^2 \| \overline{C}_j \|$. 

While the $p$ tuning parameters $\mu_j$ can be chosen by cross-validation, the computational cost explodes with $p$ to search over a grid in $p$ dimensions. Exploiting the presence of an optimal initial estimate in the form 
of $\overline{C}$, we recommend default choices for the hyperparameters as $\widehat{\mu}_j=1/\| \overline{C}_j\|^{-2}$ which in spirit is similar to the adaptive lasso \citep{zou2006adaptive}. When predictor $j$ is not important, the minimax $\ell_2$-risk for estimating $C_0^{(j)}$ is $(\log q)/n$, so that $\| \overline{C}^{(j)}\| \asymp (\log q)/n$. Since $\|X_j\|^2 \asymp n$ by assumption, see section 6, $\widehat{\mu}_j/\| X_j^\T R_j \| \asymp n^{1/2}/(\log q)^{3/2} \gg 1$, implying a strong penalty for all irrelevant predictors.

Following \cite{hahn2015decoupling}, posterior uncertainty in variable selection can be gauged if necessary by replacing $\overline{C}$ with the individual posterior samples for $C$ in \eqref{eq:utility12}.

\subsection{Post processing for rank estimation}\label{subsec:rank}
To estimate the rank, we threshold the singular values of $X \widehat{C}_R$, with $\widehat{C}_R$ obtained from \eqref{eq:subgrad2}. In situations where row sparsity is not warranted, $\overline{C}$ can be used instead of $\widehat{C}_R$. 
For $s_1, \ldots, s_q$ the singular values of $X \widehat{C}_R$, and a threshold $\omega > 0$, define the thresholded singular values as $\nu_h = s_h \, \mathrm{I}(s_h > \omega)$ for $h = 1, \ldots, q$. We estimate the rank as the number of nonzero thresholded singular values, that is, $\widehat{r}= \sum_{h=1}^q \mathrm{I}(\nu_h > 0) = \sum_{h=1}^q \mathrm{I}(s_h>\omega)$. We use the largest singular value of $Y-X\widehat{C}_R$ as the default choice of the threshold parameter $\omega$, a natural candidate for the maximum noise level in the model.

%
%
%

\section{Simulation Results}

We performed a thorough simulation study to assess the performance of the proposed method across different settings. For all our simulation settings the sample size $n$ was fixed at $100$. We considered $3$ different $(p, q)$ combinations, $(p,q)=(500,10),(200,30),(1000,12)$. The data were generated from the model $Y = X C_0 + E$. Each row of the matrix $E$ was generated from a multivariate normal distribution with diagonal covariance matrix having diagonal entries uniformly chosen between $0.5$ and $1.75$. The columns of the design matrix $X$ were independently generated from $\Gauss(0,\Sigma_X)$. We considered two cases, $\Sigma_X=\mathrm{I}_p$, and $\Sigma_X=(\sigma_{ij}^X)$, $\sigma_{jj}^X=1$, $\sigma_{ij}^X=0.5$ for $i\neq j$. The true coefficient matrix $C_0 = B_*A_*^\T$, with $B_* \in \Re^{p \times r_0}$ and $A_* \in \Re^{r \times r_0}$, with the true rank $r_0 \in \{3, 5, 7\}$. The entries of $A_*$ were independently generated from a standard normal distribution. We generated the entries in the first $s = 10$ rows of $B_*$ independently from $\mbox{N}(0, 1)$, and the remaining $(p - s)$ rows were set equal to zero. 


As a competitor, we considered the sparse partial least squares (SPLS) approach due to \cite{chun2010sparse}. Partial least squares minimizes the least square criterion between the response $Y$ and design matrix $X$ in a projected lower dimensional space where the projection direction is chosen to preserve the correlation between $Y$ and $X$ as well as the variation in $X$. \cite{chun2010sparse} suggested adding lasso type penalties while optimizing for the projection vectors for  sparse high dimensional problems. Since SPLS returns a coefficient matrix which is both row sparse and rank reduced, we create a rank reduced matrix $\widehat{C}_{RR}$ from $\widehat{C}_R$ for a fair comparison. Recalling that $\widehat{C}_R$ has zero rows, let $\widehat{S}_R$ denote the sub-matrix corresponding to the non-zero rows of $\widehat{C}_R$. Truncate the singular value decomposition of $\widehat{S}_R$ to the first $\widehat{r}$ terms where $\hat{r}$ is as obtained in \S \ref{subsec:rank}. Insert back the zero rows corresponding to $\widehat{C}_R$ in the resulting matrix to obtain $\widehat{C}_{RR}$. Clearly, $\widehat{C}_{RR} \in \Re^{p \times q}$ so created is row sparse and has rank at most $\widehat{r}$; we shall refer to $\widehat{C}_{RR}$ as the Bayesian sparse multi-task learner (BSML). 
 

For an estimator $\widehat{C}$ of $C$, we consider the mean square error, $\mathrm{MSE}=\| \widehat{C} - C_0 \|_F^2/(pq)$, and the mean square prediction error, $\mathrm{MSPE}=\| X\widehat{C} -XC_0 \|_F^2/(nq)$ to measure its performance. The squared estimation and prediction errors of SPLS and $\widehat{C}_{RR}$ for different settings are reported in table \ref{tab:mse} along with the estimates of rank. In our simulations we used the default 10 fold cross validation in the \texttt{cv.spls} function from the \texttt{R} package \texttt{spls}. The SPLS estimator of the rank is the one for which the minimum cross validation error is achieved. We observed highly accurate estimates of the rank for the proposed method, whereas SPLS overestimated the rank in all the settings considered. The proposed method also achieved superior performance in terms of the two squared errors, improving upon SPLS by as much as 5 times in some cases. Additionally, we observed that the performance of SPLS deteriorated relative to BSML with increasing number of covariates. 

\begin{figure}
\centering
\caption{Average sensitivity and specificity across 50 replicates is plotted for different choices of the postulated rank. Here $(p,q,r_0)=(1000,12,3)$. Values for BSML (SPLS) are in bold (dashed).  }

\includegraphics[width=0.95\textwidth]{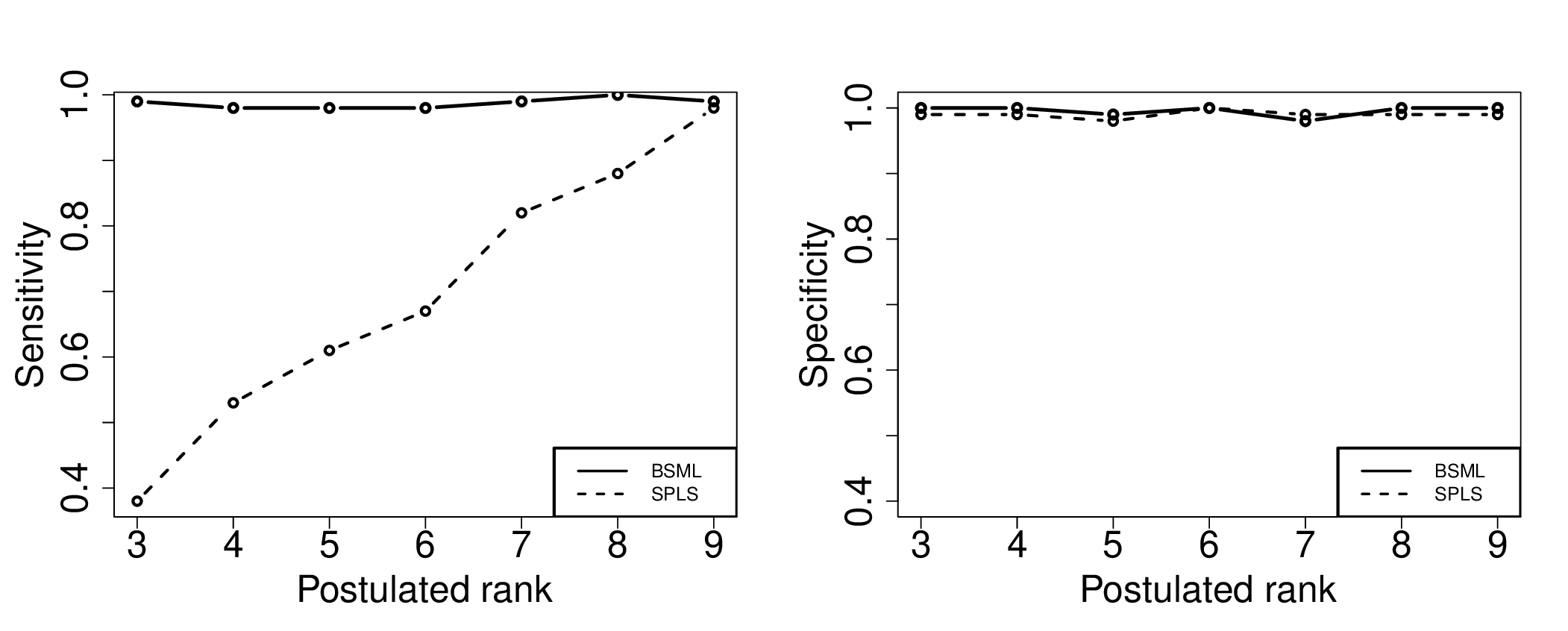}
\label{fig:sens_spec}
\end{figure}

In terms of variable selection, both methods had specificity and sensitivity both close to one in all the simulation settings listed in table \ref{tab:mse}. Since SPLS consistently overestimated the rank, we further investigated the effect of the rank on variable selection. We focused on the simulation case $(p, q, r_0) = (1000, 12, 3)$, and fit both methods with different choices of the postulated rank between 3 and 9. For the proposed method, we set $q^\ast$ in \S \ref{sec:mod} to be the postulated rank, that is, we considered $B \in \Re^{p \times q^*}$ and $A \in \Re^{q \times q^\ast}$ for $q^\ast \in \{3, \ldots, 9\}$. For SPLS, we simply input $q^\ast$ as the number of hidden components inside the function \texttt{spls}. Figure \ref{fig:sens_spec} plots the sensitivity and specificity of BSML and SPLS as a function of the postulated rank. While the specificity is robust for either method, the sensitivity of SPLS turned out to be highly dependent on the rank. The left panel of figure \ref{fig:sens_spec} reveals that at the true rank, SPLS only identifies $40 \%$ of the significant variables, and only achieves a similar sensitivity as BSML when the postulated rank is substantially overfitted. BSML, on the other hand, exhibits a decoupling effect wherein the overfitting of the rank does not impact the variable selection performance.

We conclude this section with a simulation experiment carried out in a correlated response setting. Keeping the true rank $r_0$ fixed at 3, the data were generated similarly as before except that the individual rows $e_i$ of the matrix $E$ was generated from $\Gauss(0,\Sigma)$, with $\Sigma_{ii} = 1, \Sigma_{ij} = 0.5, 1\le i \ne j \le q$. 
To accommodate the non-diagonal error covariance, we placed a inverse-Wishart$(q,\mathrm{I}_q)$ prior on $\Sigma$. 
An associate editor pointed out the recent article \citep{ruffieux2017efficient} which used spike-slab priors on the coefficients in a multiple response regression setting. They implemented a variational algorithm to posterior inclusion probabilities of each covariate, which is available from the \texttt{R} package \texttt{locus}. To select a model using the posterior inclusion probabilities, we used the median probability model \citep{barbieri2004optimal}; predictors with a posterior inclusion probability less than 0.5 were deemed irrelevant. We implemented their procedure with the prior average number of predictors to be included in the model conservatively set to 25, a fairly well-chosen value in this context. We observed a fair degree of sensitivity to this parameter in estimating the sparsity of the model, which when set to the true value $10$, resulted in comparatively poor performance whereas a value of $100$ resulted in much better performance. Table \ref{tab:var_sel} reports sensitivity and specificity of this procedure and ours, averaged over 50 replicates. While the two methods performed almost identically in the relatively low dimensional setting $(p,q)=(200,30)$, BSML consistently outperformed \cite{ruffieux2017efficient} when the dimension was higher.


\begin{center}
\begin{table}[!h]
\caption{Estimation and predictive performance of the proposed method (BSML) versus SPLS across different simulation settings. We report the average estimated rank ($\hat{r}$), Mean Square Error, MSE ($\times 10^{-4}$) and Mean Square Predictive Error, MSPE, across 50 replications. For each setting the true number of signals were 10 and sample size was 100. For each combination of $(p,q,r_0)$ the columns of the design matrix were generated from $\Gauss(0,\Sigma_X)$. Two different choices of $\Sigma_X$ was considered. $\Sigma_X=\mathrm{I}_p$ (independent) and $\Sigma_X=(\sigma_{ij}^X)$,$\sigma_{jj}^X=1$,$\sigma_{ij}^X=0.5$ for $i\neq j$ (correlated). The method achieving superior performance for each setting is highlighted in bold.}
\huge
\centering
\scalebox{0.43}{
\begin{tabular}{cccccccccccccc}\toprule
 &  & \multicolumn{12}{c}{{\bf (p,q)}}\\
 \cmidrule{3-14}
&  & \multicolumn{4}{c}{{\bf (200,30)}} & \multicolumn{4}{c}{{\bf (500,10)}} & \multicolumn{4}{c}{{\bf (1000,12)}}\\
 \cmidrule{3-14}
 &  & \multicolumn{2}{c}{\bf Independent} & \multicolumn{2}{c}{\bf Correlated} & \multicolumn{2}{c}{\bf Independent} & \multicolumn{2}{c}{\bf Correlated} & \multicolumn{2}{c}{\bf Independent} & \multicolumn{2}{c}{\bf Correlated}\\
 \cmidrule{3-14}
{{\bf Rank}} & \hspace{0.2in}{{\bf Measures}} & BSML & SPLS & BSML & SPLS & BSML & SPLS & BSML & SPLS & BSML & SPLS & BSML & SPLS \\
\cmidrule(lr){1-14}  
& $\hat{r}$ & ${\bf 3.0}$ & $7.9$ & ${\bf 3.0}$  & $9.4$ & ${\bf 3.0}$ & $9.7$ & ${\bf 3.0}$ & $8.8$ & ${\bf 3.2}$ & $9.4$ & ${\bf 3.4}$ & $8.9$ \\
{\bf 3} & $\text{MSE}$ & ${\bf 3}$ & $14$ & ${\bf 5}$ & $15$ & ${\bf 3}$ & $7$ & ${\bf 5}$ & $30$ &  ${\bf 3}$ & $50$ & ${\bf 3}$ & $38$ \\
& $\text{MSPE}$ & ${\bf 0.07}$ & $0.25$ & ${\bf 0.06}$ & ${\bf 0.17}$ & $0.22$ & ${\bf 0.15}$ & $0.34$ & $0.21$ & ${\bf 0.35}$ & $4.19$ & ${\bf 0.30}$ & $1.51$\\
\cmidrule{1-14}\\
& $\hat{r}$ & ${\bf 5}$ & $9.7$ &  ${\bf 4.9}$ & $12.2$ & ${\bf 4.9}$ & $9.9$ & ${\bf 4.8}$ & $9.8$  & ${\bf 5.1}$ & $9.9$ & ${\bf 5.1}$ & $9.9$\\
{\bf 5}& $\text{MSE}$ & ${\bf 5}$ & $69$ & ${\bf 9}$ & $61$ & ${\bf 3}$ & $10$ & ${\bf 6}$ & $24$ & ${\bf 2}$ & $108$ & ${\bf 4}$ & $129$\\
& $\text{MSPE}$ & ${\bf 0.11}$ & $3.8$ & ${\bf 0.09}$ & $4.6$ & ${\bf 0.17}$ & $0.41$ & ${\bf 0.20}$ & $0.38$ & ${\bf 0.32}$ & $9.54$ & ${\bf 0.32}$ & $4.63$\\
\cmidrule{1-14}\\
& $\hat{r}$ & ${\bf 6.9}$ & $10.3$ & ${\bf 6.9}$ & $15.8$ & ${\bf 6.8}$ & $10$ & ${\bf 6.7}$ & $9.7$ & ${\bf 6.8}$ & $10.2$ & ${\bf 6.6}$ & $11.5$\\
{\bf 7} & $\text{MSE}$ & ${\bf 6}$ & $116$ & ${\bf 10}$ & $112$ & ${\bf 3}$ & $20$ & ${\bf 5}$ & $49$ & ${\bf 2}$ & $195$ & ${\bf 4}$ & $261$ \\
& $\text{MSPE} $ & ${\bf 0.12}$ & $10.81$ & ${\bf 0.11}$ & $9.01$ & ${\bf 0.16}$ & $0.72$ & ${\bf 0.16}$ & $0.92$ & ${\bf 0.32}$ & $16.70$ & ${\bf 0.31}$ & $7.44$ \\
 \hline
 \end{tabular}
}
\label{tab:mse}
\end{table}
\end{center}
\begin{center}
\begin{table}[!h]
\caption{Variable selection performance of the proposed method in a non-diagonal error structure setting with independent and correlated predictors; $e_i\sim \Sigma$, $\sigma_{ii}=1,\sigma_{ij}=0.5$. Sensitivity and specificity of BSML is compared with \cite{ruffieux2017efficient}.}
\scalebox{0.88}{
\begin{tabular}{ccccccc}\toprule
&  & & \multicolumn{2}{c}{BSML} & \multicolumn{2}{c}{\cite{ruffieux2017efficient}}\\
\cmidrule{1-7}\\
&  ${\bf (p,q)}$ & {\bf Measure}& {\bf Independent} & {\bf Correlated} & {\bf Independent} & {\bf Correlated}\\
 \cmidrule{1-7}\\
 \multirow{9}{*}{${\bf r_0=3}$} & {\bf \multirow{2}{*}{(200,30)}} & Sensitivity & 1 & 1 & 0.96 & 0.87  \\
  & & Specificity & 0.90 & 0.84 & 0.77 & 0.67\\
  \cmidrule{1-7}\\
 & {\bf \multirow{2}{*}{(500,10)}} &   Sensitivity & 1 & 0.99 & 0.9 & 0.8\\
   & & Specificity & 0.99 & 0.99 & 0.80 & 0.64\\
   \cmidrule{1-7}\\
 & {\bf \multirow{2}{*}{(1000,12)}} &   Sensitivity & 0.99 & 0.99 & 0.92 & 0.63\\
   & & Specificity & 0.99 & 0.99 & 0.80 & 0.64\\ 
   \hline  
\end{tabular}
}
\label{tab:var_sel}
\end{table}
\end{center}


\vspace{-0.8in}
\section{Yeast Cell Cycle Data}
Identifying transcription factors which are responsible for cell cycle regulation is an important scientific problem \citep{chun2010sparse}. The yeast cell cycle data from  \cite{spellman1998comprehensive} contains information from three different experiments on mRNA levels of 800 genes on an $\alpha$-factor based experiment. The response variable is the amount of transcription (mRNA) which was measured every 7 minutes in a period of 119 minutes, a total of 18 measurements $(Y)$ covering two cell cycle periods. The ChIP-chip data from \cite{lee2002transcriptional} on chromatin immunoprecipitation contains the binding information of the 800 genes for 106 transcription factors $(X)$. We analyze this data available publicly from the \texttt{R} package \texttt{spls} which has the above information completed for 542 genes. The yeast cell cycle data was also analyzed in \cite{chen2012sparse} via sparse reduced rank regression (SRRR). Scientifically 21 transcription factors of the 106 were verified by \cite{wang2007group} to be responsible for cell cycle regulation.

\begin{figure}
\centering
\includegraphics[width=0.95\textwidth]{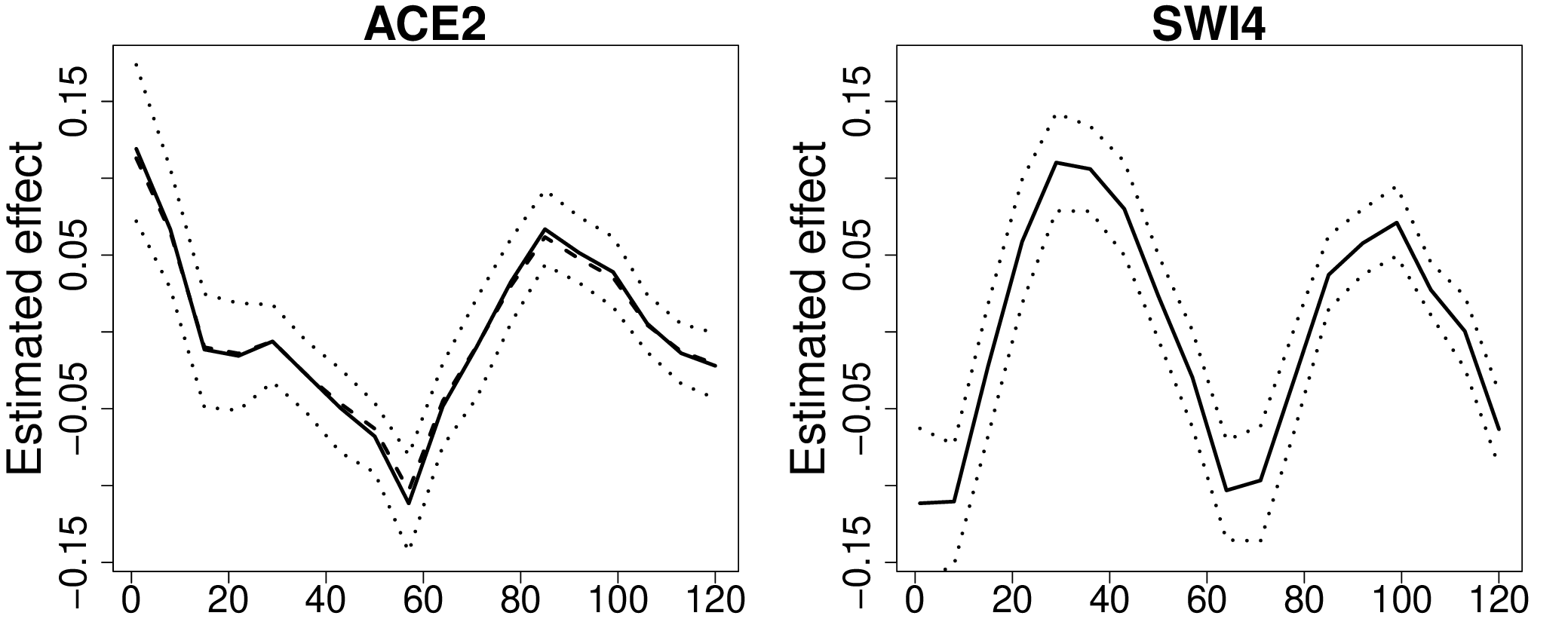}
\caption{Estimated effects of ACE2 and SWI4, two of 33 transcription factors with non-zero effects on cell cycle regulation. Both have been scientifically verified by \cite{wang2007group}. Dotted lines correspond to 95\% posterior symmetric credible intervals, bold lines represent the posterior mean and the dashed lines plot values of the BSML estimate $\widehat{C}_{RR}$.  }
\label{fig:yeast1}
\end{figure}

The proposed BSML procedure identified 33 transcription factors. Corresponding numbers for SPLS and SRRR were 48 and 69 respectively. Of the 21 verified transcription factors, the proposed method selected 14, whereas SPLS and SRRR selected 14 and 16 respectively.
10 additional transcription factors that regulate cell cycle were identified by \cite{lee2002transcriptional}, out of which 3 transcription factors were selected by our proposed method. 
Figure \ref{fig:yeast1} plots the posterior mean, BSML estimate $\widehat{C}_{RR}$, and 95 \% symmetric pointwise credible intervals for two common effects ACE2 and SW14 which are identified by all the methods. 
Similar periodic pattern of the estimated effects are observed as well for all the other two methods in contention, perhaps unsurprisingly due to the two cell cycles during which the mRNA measurements were taken. Similar plots for the remaining 19 effects identified by our method are placed inside the supplemental document. 

The proposed automatic rank detection technique estimated a rank of 1 which is significantly different from SRRR $(4)$ and SPLS $(8)$. The singular values of $Y-X\widehat{C}_R$ showed a significant drop in magnitude after the first four values which agrees with the findings in \cite{chen2012sparse}. The 10-fold cross validation error with a postulated rank of 4 for BSML was 0.009 and that of SPLS was 0.19. 

We repeated the entire analysis with a non-diagonal $\Sigma$, which was assigned an inverse-Wishart prior. No changes in the identification of transcription factors or rank estimation were detected.


\section{Concentration results}\label{sec:theory}
In this section, we establish a minimax posterior concentration result under the prediction loss when the number of covariates are allowed to grow sub-exponentially in $n$. To the best of our knowledge, this is the first such result in Bayesian reduced rank regression models. We are also not aware of a similar result involving the horseshoe or another polynomial tailed shrinkage prior in ultrahigh-dimensional settings beyond the generalized linear model framework. \cite{armagan2013posterior} applied the general theory of posterior consistency \citep{ghosal2000} to linear models with growing number of covariates and established consistency for the horseshoe prior with a sample size dependent hyperparameter choice when $p = o(n)$. Results \citep{van2014horseshoe,ghosh2017asymptotic} that quantify rates of convergence focus {\em exclusively} on the normal means problem, with their proofs crucially exploiting an exact conjugate representation of the posterior mean. 

A key ingredient of our theory is a novel non-asymptotic prior concentration bound for the horseshoe prior around sparse vectors. The prior concentration or local Bayes complexity \citep{ghosal2000,bhattacharya2016bayesian} is a key component in the general theory of posterior concentration. Let $\ell_0[s; p] = \{\theta_0 \in \Re^p: \#(1\leq j < p : \theta_{0j} \neq 0)\leq s\}$ denote the space of $p$-dimensional vectors with at most $s$ non-zero entries. 
\begin{lemma}\label{lm:lm1}
Let $\Pi_{\mathrm{HS}}$ denote the horseshoe prior on $\Re^p$ given by the hierarchy $\theta_j\mid \lambda_j,\tau \sim \Gauss(0,\lambda_j^2\tau^2),\, \lambda_j \sim \mathrm{Ca}_+(0,1),\, \tau \sim \mathrm{Ca}_+(0,1)$. Fix $\theta_0\in \ell_0[s;p]$ and let $S=\{j: \theta_{0j}\neq 0\}$. Assume $s=o(p)$ and $\log p\leq n^\gamma$ for some $\gamma \in (0,1)$ and $\max\mid \theta_{0j}\mid \leq M$ for some $M>0$ for $j \in S$. 
Define $\delta=\{(s\log p)/n\}^{1/2}$. Then,
$$\Pi_\mathrm{HS}\big(\theta\,:\, \|\theta-\theta_0\|_2<\delta \big)\geq e^{-Ks\log p},$$
for some positive constant $K$.
\end{lemma} 
A proof of the result is provided in the supplementary document. We believe Lemma \ref{lm:lm1} will be of independent interest in various other models involving the horseshoe prior, for example, high dimensional regression and factor models. 
The only other instance of a similar prior concentration result for a continuous shrinkage prior in $p \gg n$ settings that we are aware of is for the Dirichlet--Laplace prior \citep{pati2014posterior}. 

We now study concentration properties of the posterior distribution in model \eqref{eq:c_factor} in $p \gg n$ settings. 
To aid the theoretical analysis, we adopt the fractional posterior framework of \cite{bhattacharya2016bayesian}, where a fractional power of the likelihood function is combined with a prior using the usual Bayes formula to arrive at a fractional posterior distribution. Specifically, fix $\alpha \in (0,1)$ and recall the prior $\Pi_C$ on $C$ defined after equation \eqref{eq:prior} and set $\Pi_\Sigma$ as the inverse-Wishart prior for $\Sigma$. The $\alpha$-fractional posterior for $(C,\Sigma)$ under model \eqref{eq:c_factor} is then given by 
\begin{align}\label{eq:frac_post}
\Pi_{n, \alpha}(C, \Sigma \mid Y) \propto\{ p^{(n)}(Y \mid C, \Sigma; X)\}^\alpha \, \Pi_C(C) \, \Pi_\Sigma(\Sigma).
\end{align}
Assuming the data is generated with a true coefficient matrix $C_0$ and a true covariance matrix $\Sigma_0$, we now study the frequentist concentration properties of 
$\Pi_{n, \alpha}(\cdot \mid Y)$ around $(C_0, \Sigma_0)$. The adoption of the fractional framework is primarily for technical convenience; refer to the supplemental document for a detailed discussion. We additionally discuss the closeness of the fractional posterior to the usual posterior in the next subsection. 


We first list our assumptions on the truth. 
\begin{assumption}[Growth of number of covariates]
$\log p/n^\gamma \le 1$ for some $\gamma \in (0,1)$. 
\end{assumption}
\begin{assumption} The number of response variables $q$ is fixed.
\begin{assumption}[True coefficient matrix] The true coefficient matrix $C_0$ admits the decomposition $C_0=B_0A_0^\T$ where $B_0 \in \Re^{p\times r_0}$ and $A_0 \in \Re^{q \times r_0}$ for some $r_0 = \kappa q$, $\kappa\in \{1/q,2/q,\ldots, 1\}$. We additionally assume that $A_0$ is semi-orthogonal, i.e. $A_0^\T A_0= \mathrm{I}_{r_0}$, and all but $s$ rows of $B_0$ are identically zero for some $s=o(p)$. 
Finally, $\underset{j,h}{\max}\mid C_{0jh}\mid <T$ for some $T > 0$. 
\end{assumption}
\begin{assumption}[Response covariance]
The covariance matrix $\Sigma_0$ satisfies for some $a_1$ and $a_2$, $0<a_1< s_{\min}(\Sigma_0)< s_{\max}(\Sigma_0)<a_2 < \infty$ where $s_{\min}(P)$ and $s_{\max}(P)$ are the minimum and maximum singular values of a matrix $P$ respectively.
\end{assumption}

\end{assumption}
\begin{assumption}[Design matrix]
For $X_j$ the $j$th column of $X$, $\max_{1 \le j \le p} \norm{X_j} \asymp n$. 
\end{assumption}

Assumption 1 allows the number of covariates $p$ to grow at a sub-exponential rate of $e^{n^{\gamma}}$ for some $\gamma \in (0, 1)$. 
Assumption 2 can be relaxed to let $q$ grow slowly with $n$.
Assumption 3 posits that the true coefficient matrix $C_0$ admits a reduced-rank decomposition with the matrix $B_0$ row-sparse. The orthogonality assumption on true $A_0$ is made to ensure that $B_0$ and $C_0$ have the same row-sparsity \citep{chen2012sparse}. The positive definiteness of $\Sigma_0$ is ensured by assumption 4. Finally, 
assumption 4 is a standard minimal assumption on the design matrix and is satisfied with large probability if the elements of the design matrix are independently drawn from a fixed probability distribution, such as $\mbox{N}(0, 1)$ or any sub-Gaussian distribution. It also encompasses situations when the columns of $X$ are standardized. 

Let $p_0^{(n)}(Y\mid X) \equiv p^{(n)}(Y \mid C_0, \Sigma_0; X)$ denote the true density. For two densities $q_1, q_2$ with respect to a dominating measure $\mu$, recall the squared Hellinger distance $h^2(q_1,q_2)=\{(1/2)\int (q_1^{1/2} - q_2^{1/2})^2 d\mu\}$. As a loss function to measure closeness between $(C, \Sigma)$ and $(C_0, \Sigma_0)$, we consider the squared Hellinger distance $h^2$ between the corresponding densities $p(\cdot \mid C, \Sigma; X)$ and $p_0(\cdot \mid X)$. It is common to use $h^2$ to measure the closeness of the fitted density to the truth in high-dimensional settings; see, e.g., \cite{jiang2007bayesian}. In the following theorem, we provide a non-asymptotic bound to the squared Hellinger loss under the fractional posterior $\Pi_{n, \alpha}$.

\begin{theorem}\label{th:theorem1}
Suppose $\alpha \in (0,1)$ and let $\Pi_{n,\alpha}$ be defined as in \eqref{eq:frac_post}. Suppose Assumptions 1-5 are satisfied. Let the joint prior on $(C,\Sigma)$ be defined by the product prior $\Pi_C$ and $\Pi_\Sigma$ where $\Pi_\Sigma$ is the inverse-Wishart prior with parameters $(q,\mathrm{I}_q)$. Define $\widetilde{\epsilon}_n=\max\{K_1\log \rho/s_{\min}^2(\Sigma_0),4/s_{\min}^2(\Sigma_0)\}\epsilon_n$ where $\rho = s_{\max}(\Sigma_0)/s_{\min}(\Sigma_0)$, $K_1$ is an absolute positive constant, 
and $\epsilon_n=\{(qr_0+r_0s\log p)/n\}^{1/2}$. Then for for any $D\geq 1$ and $t>0$,
$$\Pi_{n,\alpha}\left[(C,\Sigma): \, h^2\big\{p^{(n)}(Y\mid C,\Sigma; \, X), \, p_0^{(n)}(Y\mid X) \big \}\geq \dfrac{(D+3t)}{2(1-\alpha)} \,n \widetilde{\epsilon}_n^{\,2}\mid Y\right]\leq e^{-tn\widetilde{\epsilon}_n^{\,2}}$$
with $P_{(C_0,\Sigma_0)}^{(n)}$ probability at least $1-K_2/\{(D-1+t)n\widetilde{\epsilon}_n^{\,2}\}$ for sufficiently large $n$ and some positive constant $K_2$.
\end{theorem} 
The proof of Theorem \ref{th:theorem1}, provided in the Appendix, hinges upon establishing sufficient prior concentration around $C_0$ and $\Sigma_0$ for our choices of $\Pi_C$ and $\Pi_\Sigma$ which in turn drives the concentration of the fractional posterior. Specifically, building upon Lemma \ref{lm:lm1} we prove in Lemma S5 of the supplementary document that for our choice of $\Pi_C$ we have sufficient prior concentration around row and rank sparse matrices. 

\cite{bunea2012joint} obtained $n\epsilon_n^2=(qr_0+r_0s\log p)$ as the minimax risk under the loss $\mid \mid XC -XC_0 \mid \mid_F^2$ for model \eqref{eq:lin_reg} with $\Sigma=\mathrm{I}_q$ and when $C_0$ satisfies assumption 3. Theorem \ref{th:theorem1} can then be viewed as a more general result with unknown covariance. Indeed, if $\Sigma=\mathrm{I}_q$, we recover the minimax rate $\epsilon_n$ as the rate of contraction of fractional posterior as stated in the following theorem. Furthermore, we show that the fractional posterior mean as a point estimator is rate optimal in the minimax sense.  For a given $\alpha \in (0,1)$ and $\Sigma=\mathrm{I}_q$, the fractional posterior simplifies to $\Pi_{n,\alpha}(C\mid Y)\propto \{p(Y \mid C, \mathrm{I}_q; X)\}^\alpha \Pi_C$.
\begin{theorem}\label{th:theorem2}
Fix $\alpha \in (0,1)$. Suppose Assumptions 1-5 are satisfied and assume that $\Sigma$ is known, and without loss of generality, equals $\mathrm{I}_q$. Let $\epsilon_n$ be defined as in Theorem \ref{th:theorem1}. Then for any $D\geq 2$ and $t>0$,
$$\Pi_{n,\alpha}\left\lbrace C\in \Re^{p\times q}: \frac{1}{nq}\| XC-XC_0\|_F^2\geq \dfrac{2(D+3t)}{\alpha(1-\alpha)}\epsilon_n^2 \mid Y\right\rbrace\leq e^{-tn\epsilon_n^2}$$
holds with $P_{C_0}^{(n)}$probability at least $1-2/\{(D-1+t)n\epsilon_n^2\}$ for sufficiently large $n$. Moreover, if $\overline{C}=\int C \Pi_{n,\alpha}(dC)$, then with $P_{C_0}^{(n)}$ probability at least $1 - K_1/\{n\epsilon_n^2\}$
$$\| X\overline{C} - XC_0 \|_F^2\, \leq \, K_2\, (qr_0+r_0 s\log p),$$
for some positive constants $K_1$ and $K_2$ independent of $\alpha$.
\end{theorem} 
The proof of Theorem \ref{th:theorem2} is provided in the Appendix. The optimal constant multiple of $\epsilon_n^2$ is attained at $\alpha=1/2$. This is consistent with the optimality of the half-power in  \cite{leung2006information} in the context of a pseudo-likelihood approach for model aggregation of least squares estimates, which shares a Bayesian interpretation as a fractional posterior. 
\subsection{Fractional and standard posteriors}\label{sec:fracVSstd}

From a computational point of view, for model \eqref{eq:lin_reg}, raising the likelihood to a fractional power only results in a change in the (co)variance term, and hence our Gibbs sampler discussed subsequently can be easily adapted to sample from the fractional posterior. We conducted numerous simulations with values of $\alpha$ close to 1 and obtained virtually indistinguishable point estimates  compared to the full posterior; details are provided in the supplemental  document. In this subsection we study the closeness between the fractional posterior $\Pi_{n,\alpha}(\cdot \mid Y)$ and the standard posterior $\Pi_n(\cdot \mid Y)$ for model \eqref{lm:lm1} with prior $\Pi_C \otimes \Pi_\Sigma$ in terms of the total variation metric. Proofs of the results in this section are collected in the supplementary document.

Recall that for two densities $g_1$ and $g_2$ with respect to some measure $\mu$, the total variation distance between them is given by $\norm{g_1 - g_2}_{\mathrm{TV}} = \int \vert g_1 - g_2 \vert d\mu = \sup_{B \in \mathcal{B}} \vert G_1(B) - G_2(B) \vert$, where $G_1$ and $G_2$ denote the corresponding probability measures.
\begin{theorem}\label{th:tv_theorem}
Consider model \eqref{eq:lin_reg} with $C \sim \Pi_C$, and $ \Sigma \sim \Pi_\Sigma.$ Then,
$$ \lim_{\alpha \to 1_{-}}\norm {\Pi_{n,\alpha}(C,\Sigma \mid Y) - \Pi_{n}(C,\Sigma \mid Y)}_{\mathrm{TV}} = 0, $$
for every $Y \sim P_{C_0}$.
\end{theorem}
\cite{bhattacharya2016bayesian} proved a weak convergence result under a more general setup whereas Theorem \ref{th:tv_theorem} provides a substantial improvement to show strong convergence for the Gaussian likelihood function considered here. The total variation distance is commonly used in Bayesian asymptotics to justify posterior merging of opinion, i.e., the total variation distance between two posterior distributions arising from two different priors vanish as sample size increases. Theorem \ref{th:tv_theorem} has a similar flavor, with the exception that the merging of opinion takes place under small perturbations of the likelihood function. 

We conclude this section by showing that the regular posterior $\Pi_{n}(C, \Sigma \mid Y)$ is consistent, leveraging on the contraction of the fractional posteriors $\Pi_{n,\alpha}(C, \Sigma\mid Y)$ for any $\alpha < 1$ in combination with Theorem \ref{th:tv_theorem} above. 
For ease of exposition we assume $\Sigma = \mbox{diag}(\sigma_1^2, \ldots, \sigma_q^2)$ and $\Pi_\Sigma(\cdot)$ is a product prior with components $\mbox{inverse-Gamma}(a,b)$ for some $a, b>0$. Similar arguments can be made for the inverse-Wishart prior when $\Sigma$ is non-diagonal.
\begin{theorem}\label{thm:cons}
Assume $\Sigma = \mbox{diag}(\sigma_1^2, \ldots, \sigma_q^2)$ in model \eqref{eq:lin_reg} with priors $\Pi_C$ and a product $\mbox{inverse-Gamma}(a,b)$ prior on $\Sigma$ with $a,b>0$. For any $\epsilon > 0$ and sufficiently large $M$, 
$$
\lim_{n \to \infty} \Pi_n\bigg[ (C,\Sigma): \frac{1}{n} \, h^2\big\{p^{(n)}( \cdot \mid C,\Sigma; \, X), \, p_0^{(n)}(\cdot \mid X) \big \}\geq  M \epsilon \mid Y \bigg] \to 0,
$$
almost surely under $P_{(C_0,\Sigma_0)}$.
\end{theorem}
Theorem \ref{thm:cons} establishes consistency of the regular posterior under the average Hellinger metric $n^{-1} h^2\big\{p^{(n)}( \cdot \mid C,\Sigma; \, X), \, p_0^{(n)}(\cdot \mid X) \big \}$. This is proved using a novel representation of the regular posterior as a fractional posterior under a different prior, a trick which we believe will be useful to arrive at similar consistency results in various other high-dimensional Gaussian models. 
For any $\alpha \in (0,1)$, 
\begin{align*}
\Pi_{n}(C, \Sigma \mid Y)
& \propto |\Sigma|^{-n/2} \, e^{- \mbox{tr}\{ (Y - X C)\Sigma^{-1} (Y - X C)^{\T} \}/2} \, \Pi_C(dC) \Pi_\Sigma(d\Sigma)\\
& \propto |\Sigma|^{-n \alpha/2} \, e^{- \alpha \mbox{tr}\{ (Y - X C)(\alpha \Sigma)^{-1} (Y - X C)^{\T} \}/2}  \, \Pi_C(dC) \, |\Sigma|^{-n (1-\alpha)/2} \,  \Pi_\Sigma(d\Sigma) \\
& \propto |\Sigma_*|^{-n\alpha/2} \, e^{- \alpha \mbox{tr}\{ (Y - X C)\Sigma_*^{-1} (Y - X C)^{\T} \}/2} \, \Pi_C(dC) \Pi_{\Sigma_*}(d\Sigma_*)\\
& \propto \Pi_{n,\alpha}(C, \Sigma_* \mid Y),
\end{align*} 
where $\Sigma_* = \alpha \Sigma$ and from a simple change of variable, $\Pi_{\Sigma_*}(\cdot)$ is again a product of inverse-Gamma densities with each component a $\mbox{inverse-Gamma}\{n(1-\alpha)/2 + a, \alpha b\}$. Since the first and last expressions in the above display are both probability densities, we conclude that $\Pi_{n}(C, \Sigma \mid Y) = \Pi_{n,\alpha}(C, \Sigma_* \mid Y)$. This means that the regular posterior distribution of $(C, \Sigma)$ can be viewed as the $\alpha$-fractional posterior distribution of $(C, \Sigma_*)$, with the prior distribution of $\Sigma_*$ dependent on both $n$ and $\alpha$. Following an argument similar to Theorem \ref{th:theorem1}, we only need to show the prior concentration of $(C, \Sigma_*)$ around the truth to obtain posterior consistency of $\Pi_{n,\alpha}(C, \Sigma_* \mid Y)$, and hence equivalently of $\Pi_{n}(C, \Sigma \mid Y)$. The only place that needs additional care is showing the prior concentration of $\Sigma_*$ with an $n$-dependent prior. This can be handled by setting $\alpha = 1 - 1/(\log n)^t$ for some appropriately chosen $t > 1$.

\appendix

\section*{Appendix}
Lemmas numbered as S1, S2 etc. refer to technical lemmas included in the supplementary material. For two densities $p_\theta$ and $p_{\theta_0}$ with respect to a common dominating measure $\mu$ and indexed by parameters $\theta$ and $\theta_0$ respectively, the R\'enyi divergence of order $\alpha \in (0,1)$ is defined as $D_\alpha(\theta,\theta_0)=(\alpha-1)^{-1}\log \int p_\theta^\alpha p_{\theta_0}^{1-\alpha}d\mu$. The $\alpha$-affinity between $p_\theta$ and $p_{\theta_0} $ is denoted by $A_\alpha(p_\theta,p_{\theta_0})=\int p_\theta^\alpha p_{\theta_0}^{1-\alpha}d\mu= e^{-(1-\alpha)D_\alpha(p_\theta,p_{\theta_0})}$. See \cite{bhattacharya2016bayesian} for a review of R\'enyi divergences. 
\subsection*{Proof of theorem \ref{th:theorem1}}
\begin{proof}
Fix $\alpha \in [1/2,1)$. Define $U_n=\left[(C,\Sigma):\dfrac{1}{n}D_\alpha\{(C,\Sigma),(C_0,\Sigma_0)\}>\dfrac{D+3t}{1-\alpha}\tilde{\epsilon_n}^2\right]$. Let $\eta=(C,\Sigma)$ and $\eta_0=(C_0,\Sigma_0)$. Also let $p_\eta^{(n)}$ denote the density of $Y\mid X$ with parameter value $\eta$ under model (\ref{eq:c_factor}). Finally, let $\Pi_\eta$ denote the joint prior $\Pi_C \times \Pi_\Sigma$. Then, the $\alpha$-fractional posterior probability assigned to the set $U_n$ can be written as,
\begin{equation}
\Pi_{n,\alpha}(U_n\mid Y)=\dfrac{\int_{U_n}e^{-\alpha r_n(\eta,\eta_0)}d\Pi_\eta}{\int e^{-\alpha r_n(\eta,\eta_0)}d\Pi_\eta}:= \frac{N_n}{D_n},
\end{equation}
where $r_n(\eta,\eta_0)=\log p_{\eta_0}^{(n)}/p_\eta^{(n)}$. We prove in lemma S6 of the supplementary material that, with $P_{(C_0,\Sigma_0)}^{(n)}$-probability at least $1-K_2/\{(D-1+t)^2n\tilde{\epsilon}_n^2\}$
$D_n \geq e^{-\alpha(D+t)n\tilde{\epsilon}_n^2}$ for some positive constant $K_2$. For the numerator proceeding similarly as in the proof of theorem 3.2 in \cite{bhattacharya2016bayesian} we arrive at,
$P_{(C_0,\Sigma_0)}^{(n)}\left\lbrace N_n \leq e^{-(D+2t)n\tilde{\epsilon}_n^2}\right\rbrace \geq 1 - 1/\{(D-1+t)^2n\tilde{\epsilon}_n^2\}.$
Combining the upper bound for $N_n$ and lower bound for $D_n$ we then have,
$$\Pi_{n,\alpha}\left[(C,\Sigma):\dfrac{1}{n}D_\alpha\{(C,\Sigma),(C_0,\Sigma_0)\}\geq \dfrac{(D+3t)}{1-\alpha}\epsilon_n^2\mid Y\right]\leq e^{-tn\epsilon_n^2},$$
with $P_{\eta_0}^{(n)}$-probability at least $1-K_2/\{(D-1+t)^2n\tilde{\epsilon}_n^2\}$.
Since by assumption $\alpha\geq 1/2$, we have the relation, $D_\alpha(p,q) \geq D_{1/2}(p,q)\geq 2 h^2 (p,q)$ \citep{bhattacharya2016bayesian} for two densities $p$ and $q$ proving the theorem.
\end{proof}

\subsection*{Proof of theorem \ref{th:theorem2}}
\begin{proof}
For $C \in \Re^{p \times q}$, we write $p_C^{(n)}$ to denote the density of $Y|X$ which is proportional to $e^{- \, \mbox{tr}\{ (Y - X C) (Y - X C)^{\T} \}/2 }$. For any $C^*\in \Re^{p \times q}$ we define a $\epsilon$-neighborhood as,
\begin{equation}
B_n(C^*,\epsilon)=\left\lbrace C \in \Re^{p \times q}: \int p_{C^*}^{(n)} \log (p_{C^*}^{(n)}/p_{C}^{(n)})dY\leq n\epsilon^2,\int p_{C^*}^{(n)} \log^2 (p_{C^*}^{(n)}/p_{C}^{(n)})dY\leq n\epsilon^2\right\rbrace.
\end{equation}
Observe that $B_n(C_0,\epsilon)\supset A_n(C_0,\epsilon)=\left\lbrace C\in \Re^{p\times q}:  \frac{1}{n}\| XC-XC_0\|_F^2\leq \epsilon^2\right\rbrace$ for all $\epsilon>0$ and the R\'enyi divergence $D_\alpha(p_C^{(n)},p_{C_0}^{(n)})=\frac{\alpha}{2}\| XC-XC_0\|_F^2$. By a similar argument as in step 1 of the proof of lemma S6 of the supplementary document, we have $\Pi_C\{A_n(C_0,\epsilon_n)\}\geq e^{-Kn\epsilon_n^2}$ for positive $K$. Hence the theorem follows from theorem 3.2 of \cite{bhattacharya2016bayesian}.  
\end{proof}

\bibliography{sparse_reduced_rank_refs}
\bibliographystyle{biometrika}

\newpage
\setcounter{equation}{0}
\setcounter{page}{1}
\setcounter{table}{1}
\setcounter{section}{0}
\numberwithin{table}{section}
\renewcommand{\theequation}{S.\arabic{equation}}
\renewcommand{\thesubsection}{S.\arabic{section}.\arabic{subsection}}
\renewcommand{\thesection}{S.\arabic{section}}
\renewcommand{\thetable}{S.\arabic{table}}
\renewcommand{\thefigure}{S\arabic{figure}}
\renewcommand{\thelemma}{S\arabic{lemma}}
\renewcommand{\bibnumfmt}[1]{[S#1]}
\renewcommand{\citenumfont}[1]{S#1}

\begin{center}
\Large{\bf Supplementary material}
\end{center}

\vskip 10mm
\begin{center}
Antik Chakraborty\\
Department of Statistics, Texas A\&M University, College Station\\
3143 TAMU, TX 77843-3143, USA\\
antik@stat.tamu.edu\\
\hskip 5mm\\
Anirban Bhattacharya\\
Department of Statistics, Texas A\&M University, College Station\\
3143 TAMU, TX 77843-3143, USA\\
anirbanb@stat.tamu.edu\\
\hskip 5mm\\
Bani K. Mallick\\
Department of Statistics, Texas A\&M University, College Station\\
3143 TAMU, TX 77843-3143, USA\\
bmallick@stat.tamu.edu\\

\end{center}
\section*{Convention}
Equations defined in this document are numbered (S1), (S2) etc, while (1), (2) etc refer to those defined in the main document. Similar for lemmas, theorems etc. Throughout the document we use $K$, $T$ for positive constants whose value might change from one line to the next. Then notation $a\lesssim b$ means $a\leq K b$. For an $m \times r$ matrix $A$ (with $m > r$), $s_i(A) = \sqrt{\lambda_i}$ for $i = 1, \ldots, r$ denote the singular values of $A$, where $\lambda_1 \ge \lambda_2 \ge \ldots \ge \lambda_r \ge 0$ are the eigenvalues of $A^\T A$. The largest and smallest singular values will be denoted by $s_{\max}(A)$ and $s_{\min}(A)$. The operator norm of $A$, denoted $\|A\|_2$, is the largest singular value $s_{\max}(A)$. The Frobenius norm of $A$ is $\|A\|_F = (\sum_{i=1}^m \sum_{j=1}^r a_{ij}^2)^{1/2}$. 

\section*{Fractional versus usual posterior}
In this section, we provide some additional discussion regarding our adoption of the fractional posterior framework in the main document. We begin with a detailed discussion on the sufficient conditions required to establish posterior contraction rates for the usual posterior from \cite{ghosal2000} and contrast them with those of fractional posteriors \citep{bhattacharya2016bayesian}. For simplicity, we discuss the i.i.d. case although the discussion is broadly relevant beyond the i.i.d. setup. We set out with some notation. Suppose we observe $n$ independent and identically distributed random variables $X_1,\ldots, X_n\mid P \sim P$ where $P\in \mathcal{P}$, a family of probability measures. Denote $L_n(P)$ as the likelihood for this data which we abbreviate and write as $X^{(n)}$. We treat $P$ as our parameter of interest and define a prior $\Pi_n$ for $P$. 

Let $P_0 \in \mathcal{P}$ be the true data generating distribution. For a measurable set $B$, the posterior probability assigned to $B$ is 
\begin{equation}\label{eq:def_post}
\Pi_n(B \mid X^{(n)}) =\dfrac{\int_B L_n(P) \, \Pi_n(dP)}{\int_\mathcal{P} L_n(P) \, \Pi_n(dP)}
\end{equation}

For $\alpha\in (0,1)$, the $\alpha$-fractional posterior $\Pi_{n,\alpha}(\cdot\mid Y)$ is,
\begin{equation}\label{eq:def_frac_post}
\Pi_{n,\alpha}(B \mid X^{(n)}) =\dfrac{\int_B \{L_n(P)\}^\alpha\Pi_n(dP)}{\int_\mathcal{P} \{L_n(P)\}^\alpha \Pi_n(dP)}.
\end{equation}
The fractional posterior is obtained upon raising the likelihood to a fractional power $\alpha$ and combining with the prior using Bayes's theorem. 

Let $p$ and $p_0$ be the density of $P$ and $P_0$ respectively with respect to some measure $\mu$ and $p^{(n)}$ and $p_0^{(n)}$ be the corresponding joint densities. Suppose $\epsilon_n$ is a sequence such that $\epsilon_n \rightarrow 0$ and $n \epsilon_n^2 \rightarrow \infty$ as $n \rightarrow \infty$. Define $B_n = \{p: \int p_0^{(n)} \log p_0^{(n)}/p^{(n)}\leq n\epsilon_n^2, \int p_0^{(n)} \log^2 p_0^{(n)}/p^{(n)}\leq n\epsilon_n^2\}$. Given a metric $\rho$ on $\mathcal{P}$ and $\delta>0$, let $N(P^*,\rho, \delta)$ be the covering number of $P^*\subset\mathcal{P}$ \citep{ghosal2000}. For sake of concreteness, we focus on the case where $\rho$ is the Hellinger distance.
We now state the sufficient conditions for $\Pi_n(\cdot \mid X^{(n)})$  to contract at rate $\epsilon_n$ at $P_0$ \citep{ghosal2000}.

\begin{theorem}[\cite{ghosal2000}] \label{th:ghosal}
Suppose $\epsilon_n$ be as above. If,
there exists $\mathcal{P}_n \subset \mathcal{P}$ and positive constants $C_1, C_2$ such that,

1. $\log N(\mathcal{P}_n,h,\epsilon_n)\lesssim n\epsilon_n^2$,

2. $\Pi_n(\mathcal{P}_n^c)\leq e^{-C_1 n\epsilon_n^2}$, and

3. $\Pi_n(B_n)\geq e^{-C_3 n\epsilon_n^2}$,

then $\Pi_n\{p:h^2(p,p_0)> M\epsilon_n\mid X^{(n)}\}\rightarrow 0$ in $P_0$-probability for a sufficiently large $M$.
\end{theorem}
However, if we use the fractional posterior $\Pi_{n,\alpha}(\cdot \mid X^{(n)})$ for $\alpha \in (0,1)$, then we have the following result from \cite{bhattacharya2016bayesian},

\begin{theorem}[\cite{bhattacharya2016bayesian}]\label{th:frac_post}
Suppose condition 3 from Theorem S1 is satisfied. Then $\Pi_{n,\alpha}\{h^2(p,p_0)>M\epsilon_n \mid X^{(n)}\} \rightarrow 0$ in $P_0$-probability.
\end{theorem}

We refer the reader to \cite{bhattacharya2016bayesian} for a more precise statement of Theorem \ref{th:frac_post}. The main difference between Theorems \ref{th:ghosal} and \ref{th:frac_post} is that the same rate of convergence (upto constants) can be arrived at 
verifying fewer conditions. The construction of the sets $\mathcal{P}_n$, known as sieves, can be challenging for heavy-tailed priors such as the horseshoe. On the other hand, one only needs to verify the prior concentration bound $\Pi_n(B_n)\geq e^{-C_3 n\epsilon_n^2}$ to ensure contraction of the fractional posterior. This allows one to obtain theoretical justification in complicated high-dimensional models as in ours. To quote the authors of \cite{bhattacharya2016bayesian}, `{\it the condition of exponentially decaying prior mass assigned to the
complement of the sieve implies fairly strong restrictions on the prior tails and essentially rules
out heavy-tailed prior distributions on hyperparameters. On the other hand, a much broader class
of prior choices lead to provably optimal posterior behavior for the fractional posterior}'. That said, the proof of the technical results below illustrate that verifying the prior concentration condition alone can pose a stiff technical challenge. 

We now aim to provide some intuition behind why the theory simplifies with the fractional posterior. Define $U_n=\{p: h^2(p,p_0)>M\epsilon_n\}$. From equation (R2) and (R3) in \cite{bhattacharya2016bayesian} $U_n$ can be alternatively defined as, $U_n=\{p:D_\alpha(p,p_0)>M^* \epsilon_n\}$, where the constant $M^*$ can be derived from $M$ by the equivalence relation R\'enyi divergences \cite[equation (R3)]{bhattacharya2016bayesian}. The posterior probability assigned to the set $U_n$ is then obtained by \eqref{eq:def_post} and the fractional posterior probability assigned to $U_n$ follows from \eqref{eq:def_frac_post}. Thus after dividing the numerator and denominator by the appropriate power of $L_n(P_0)$ we get,
\begin{equation}\label{eq:post_u_n}
\Pi(U_n \mid X^{(n)}) =\dfrac{\displaystyle \int_{U_n} \dfrac{L_n(P)}{L_n(P_0)}\Pi_n(dP)}{\displaystyle \int_\mathcal{P} \dfrac{L_n(P)}{L_n(P_0)}\Pi_n(dP)},
\end{equation} 
and 
\begin{equation}\label{eq:frac_post_u_n}
\Pi_{n,\alpha}(U_n \mid X^{(n)}) =\dfrac{\displaystyle\int_{U_n} \left\lbrace\dfrac{L_n(P)}{L_n(P_0)}\right\rbrace^\alpha\Pi_n(dP)}{\displaystyle\int_\mathcal{P} \left\lbrace \dfrac{L_n(P)}{L_n(P_0)}\right\rbrace^\alpha \Pi_n(dP)}.
\end{equation}
Taking expectation of the numerator in \eqref{eq:frac_post_u_n} with respect to $P_0$ and applying Fubini's theorem to interchange the integrals yields $\int_{U_n}e^{-(1-\alpha)D_\alpha(p,p_0)} \, \Pi_n(dP)$ which by definition of $U_n$ is small. The same operation for \eqref{eq:post_u_n} leads to $\int_{U_n}\Pi_n(dP)$ which isn't necessarily small, needing the introduction of the sieves $\mathcal{P}_n$ in the analysis. 


We conducted a small simulation study carried out to compare the results of $\Pi_n$ and $\Pi_{n,\alpha}$ for different choices of $\alpha$ in the context of model (3) in the main document with priors defined in (4). We obtain virtually indistinguishable operating characteristics of the point estimates, further corroborating our theoretical study.

\begin{center}
\begin{table}[!h]
\caption{Empirical results comparing $\hat{r}$, MSPE = $(nq)^{-1} \| XC -XC_0 \mid\mid_F^2 $ and MSE=$(pq)^{-1}\| C - C_0 \|_F^2$ for different choices of the fractional power $\alpha$. $\alpha=1$ corresponds to the usual posterior. The data used in this table was generated in a similar manner as described in section 4 of the main document.}
\huge
\centering
\scalebox{0.43}{
\begin{tabular}{cccccccccccccc}\toprule
 &  & \multicolumn{12}{c}{{\bf (p,q)}}\\
 \cmidrule{3-14}
&  & \multicolumn{4}{c}{{\bf (200,30)}} & \multicolumn{4}{c}{{\bf (500,10)}} & \multicolumn{4}{c}{{\bf (1000,12)}}\\
 \cmidrule{3-14}
 &  & \multicolumn{2}{c}{\bf Independent} & \multicolumn{2}{c}{\bf Correlated} & \multicolumn{2}{c}{\bf Independent} & \multicolumn{2}{c}{\bf Correlated} & \multicolumn{2}{c}{\bf Independent} & \multicolumn{2}{c}{\bf Correlated}\\
 \cmidrule{3-14}
{{\bf $\alpha$}} & \hspace{0.2in}{{\bf Measures}} & BSML & SPLS & BSML & SPLS & BSML & SPLS & BSML & SPLS & BSML & SPLS & BSML & SPLS \\
\cmidrule(lr){1-14}  
& $\hat{r}$ & $ 3.0$ & $7.9$ & $ 3.0$  & $9.4$ & $ 3.0$ & $9.7$ & $ 3.0$ & $8.8$ & $3.2$ & $9.4$ & $3.4$ & $8.9$ \\
{\bf 1} & $\text{MSE}$ & $ 3$ & $14$ & $ 5$ & $15$ & $ 3$ & $7$ & $ 5$ & $30$ &  $ 3$ & $50$ & $ 3$ & $38$ \\
& $\text{MSPE}$ & $ 0.07$ & $0.25$ & $ 0.06$ &  $0.17$ & $0.22$ & $ 0.15$ & $0.34$ & $0.21$ & $ 0.35$ & $4.19$ & $ 0.30$ & $1.51$\\
\cmidrule{1-14}\\
& $\hat{r}$ & $ 3.0$ &  &  $ 3.0$ &  & $ 3.0$ &  & $ 3.0$ &   & $ 3.0$ &  & $ 3.1$ & \\
{\bf 0.5}& $\text{MSE}$ & $ 1.9$ &  & $ 2.7$ &  & $ 1.9$ &  & $ 3.9$ &  & $ 1.2$ &  & $ 1.4$ & \\
& $\text{MSPE}$ & $ 0.05$ &  & $ 0.06$ &  & $ 0.15$ &  & $ 0.25$ &  & $ 0.22$ &  & $ 0.32$ & \\
\cmidrule{1-14}\\
& $\hat{r}$ & $ 3.1$ &  & $ 3.0$ &  & $ 3.0$ &  & $ 2.9$ &  & $ 2.9$ &  & $ 3.0$ & \\
{\bf 0.75} & $\text{MSE}$ & $ 1.8$ &  & $ 2.4$ &  & $ 1.6$ &  &$4.3$  &  & $ 1.2$ &  & $ 1.2$ & \\
& $\text{MSPE} $ & $ 0.08$ &  & $ 0.07$ &  & $ 0.16$ &  & $ 0.22$ &  & $ 0.32$ &  & $ 0.31$ &  \\
\cmidrule{1-14}\\
& $\hat{r}$ & $ 3.0$ &  & $ 3.1$ &  & $ 3.0$ &  & $ 3.0$ &  & $ 3.1$ &  & $ 2.9$ & \\
{\bf 0.95} & $\text{MSE}$ & $ 2.1$ &  & $ 2.9$ &  & $ 1.5$ &  & $ 3.6$ &  & $ 1.5$ &  & $ 1.5$ & \\
& $\text{MSPE} $ & $ 0.09$ &  & $ 0.07$ &  & $ 0.16$ & & $ 0.29$ &  & $ 0.31$ &  & $ 0.31$ &  \\
 \hline
 \end{tabular}
}
\label{tab:mse}
\end{table}
\end{center}

We end this section by recording a high probability risk bound for the R\'enyi loss \citep{bhattacharya2016bayesian} in the following corollary that is subsequently used.
\begin{corollary}[\cite{bhattacharya2016bayesian}]
Fix $\alpha \in (0,1)$. Recall the definition of R\'enyi divergence between $p$ and $p_0$ from the main document: $D_\alpha(p,p_0)=(\alpha-1)^{-1}\, \log \int p^\alpha p_0 ^{1-\alpha}d\mu$. Under the conditions of Theorem \ref{th:frac_post},	for any $k\geq 1$,
$$\int \left\lbrace \dfrac{1}{n}D_\alpha(p,p_0)\right\rbrace^k d \Pi_{n,\alpha}(\cdot \mid X^{(n)})\leq K_1 (1-\alpha)^{-k}\epsilon_n^{2k},$$
with $P_0$-probability at least $1-K_2/(n\epsilon_n^2)$, where $K_1$ and $K_2$ are positive constants independent of $\alpha$.
\end{corollary}

\section{Proof of Lemma 1}
The following lemma is a novel prior concentration result for the horseshoe prior which bounds from below the probability assigned to an Euclidean neighborhood of a sparse vector. 
\begin{proof}
Using the scale-mixture formulation of $\Pi_{\HS}$,
\begin{equation}
\begin{split}
\Pi_\HS(\| \beta-\beta_0\|_2<\delta)&=\int_\tau pr(\| \beta-\beta_0\|_2<\delta\mid \tau) f(\tau) d\tau\\
& \geq \int_{I_{\tau_*}} pr(\| \beta-\beta_0\|_2<\delta\mid \tau) f(\tau) d\tau,\\
\end{split}
\end{equation}
where $I_{\tau_*}=[\tau_*/2,\tau_*]$ with $\tau_*=\left(s/p\right)^{3/2}\{(\log p)/n\}^{1/2}$. Let $S=\{1\leq j \leq p : \theta_{0j}\neq 0\}$.
We first provide a lower bound of the conditional probability $pr(\| \beta-\beta_0\|_2<\delta\mid \tau \in I_{\tau_*})$. For $\tau \in I_{\tau^*}$, we have,
\begin{equation}\label{eq:split_prob}
\begin{split}
pr(\| \beta-\beta_0\|_2 <\delta\mid \tau)&\geq pr\left(\|\beta_S-\beta_{0S}\|_2<\delta/2 \mid \tau\right) pr\left(\| \beta_{S^c}\|_2<\delta/2 \mid \tau\right)\\
&\geq \prod_{j\in S}pr\left(|\beta_j-\beta_{0j}|<\frac{\delta}{2\sqrt{s}}\mid \tau\right) \prod_{j \in S^c}pr\left(|\beta_j|<\frac{\delta}{2\sqrt{p}}\mid \tau\right).\\
\end{split}
\end{equation}
For a fixed $\tau \in I_{\tau_*}$, we will provide lower bounds for each of the terms in the right hand side of \eqref{eq:split_prob}; $pr\{| \beta_j-\beta_{0j}| <\delta/(2s^{1/2})\}$ for any $j \in S$, and $pr\{|\beta_j| <\delta/(2p^{1/2})\}$ for any $j \in S^c$. 

We first consider $pr\{\mid \beta_j \mid<\delta /(2p^{1/2})\mid \tau\}$ with $\tau \in I_{\tau_*}$. Since given $\tau$ and $\lambda$, $\beta_j\sim \Gauss(0,\lambda_j^2\tau^2)$, we use the Chernoff bound for a Gaussian random variable to obtain,
\begin{align*}
pr\big\{ |\beta_j| > \delta/(2p^{1/2}) \mid \lambda_j, \tau\big \}\leq 2e^{-\delta^2/(8p\lambda_j^2 \tau^2)}\leq 2e^{-\delta^2/(8p\lambda_j^2 \tau_*^2)}=2e^{-p^2/(8s^2\lambda_j^2)},
\end{align*}
since $n\delta^2=s\log p$. Thus, 
\begin{align*}
pr\big\{ | \beta_j | <\delta/(2p^{1/2}) \mid \tau \big \} =&\int_{\lambda_j} pr \big\{ |\beta_j| < \delta/(2p^{1/2}) \mid \lambda_j, \tau\big\} \, f(\lambda_j) \,d\lambda_j\\
& \geq \int_{\lambda_j} \left\lbrace1-2\exp{\left(-\dfrac{p^2}{8s^2\lambda_j^2}\right)}\right\rbrace  \, f(\lambda_j) \,d\lambda_j\\
&= 1-\dfrac{4}{\pi}\int_{\lambda_j} \exp{\left(-\dfrac{p^2}{8s^2\lambda_j^2}\right)} (1+\lambda_j^2)^{-1}d\lambda_j=1-\dfrac{4}{\pi} \, \mathcal{I}, 
\end{align*}
where $\mathcal{I}=\int_{\lambda_j} \exp{\big\{-p^2/(8s^2\lambda_j^2)\big\}} \, (1+\lambda_j^2)^{-1}d\lambda_j$. We then bound the integrand from above as follows,
\begin{align*}
\mathcal{I}=\int_{\lambda_j} \exp{\left(-\dfrac{p^2}{8s^2\lambda_j^2}\right)} \, (1+\lambda_j^2)^{-1}d\lambda_j &\leq \int_{\lambda_j} \exp{\left(-\dfrac{p^2}{8s^2\lambda_j^2}\right)}\lambda_j^{-2} d\lambda_j\\
&= \frac{1}{2} \int_0^{\infty} z^{-1/2}  \exp{\left({-\dfrac{p^2z}{8s^2}}\right)}dz, \\
&=\dfrac{\Gamma (1/2)}{\{2 p^2/(8s^2)\}^{1/2}} =  \frac{s\sqrt{2\pi}}{p},
\end{align*}
where we made the substitution $z=1/\lambda^2$ at the third step. Thus, for $\tau \in I_{\tau_*}$, $pr(\mid \beta_j \mid < \delta/2p^{1/2}\mid \tau)\geq 1-R s/p$, where $R=(32/\pi)^{1/2}$. \\

Next, for $pr(\mid \beta_j-\beta_{0j} \mid <\delta_0|\tau)$ with $\tau \in I_{\tau_*}$, we have, letting $\delta_0=s^{-1/2}(\delta/2) = 2^{-1}\{(\log p)/n\}^{1/2}$,
\begin{align*}
\begin{split}
pr(\mid \beta_j-\beta_{0j}\mid <\delta_0 \mid \tau)&=(2/\pi^3)^{1/2}\int_{\lambda_j}\int_{\mid \beta_j-\beta_0 \mid <\delta_0} \exp\{{-\beta_j^2/(2\lambda_j^2\tau^2)}\} \dfrac{1}{\lambda_j\tau(1+\lambda_j^2)} \,d\beta_j d\lambda_j\\
& \geq (2/\pi^3)^{1/2}\int_{|\beta_j-\beta_0|<\delta_0} \int_{1/\tau}^{2/\tau} \exp\{{-\beta_j^2/(2\lambda_j^2\tau^2)}\} \dfrac{1}{\lambda_j\tau(1+\lambda_j^2)} \,d\lambda_j d\beta_j\\
& \geq (2/\pi^3)^{1/2} \int_{|\beta_j-\beta_0|<\delta_0} \exp(-\beta_j^2/2)  \left(\int_{1/\tau}^{2/\tau} \dfrac{1}{1+\lambda_j^2} \,d\lambda_j \right) d\beta_j,\\
\end{split}
\end{align*}
since for $\lambda_j \in [1/\tau, 2/\tau]$, $1/(\lambda_j \tau) \geq 1/2$ and $\exp\{{-\beta_j^2/(2\lambda_j^2\tau^2)}\} \geq \exp{(-\beta_j^2/2)}$. Continuing,
\begin{align*}
pr(\mid \beta_j-\beta_{0j} \mid <\delta_0\mid \tau)&\geq (2/\pi^3)^{1/2}\,\, \dfrac{\tau}{4+\tau^2}\, \int_{|\beta_j-\beta_0|<\delta_0} \exp{(-\beta_j^2/2)}d\beta_j\\
& \geq (2/\pi^3)^{1/2}\,\, \dfrac{\tau}{4+\tau^2}\,\, \exp\{{-(M+1)^2/2}\}\,\,\delta_0\\
&\geq K \, \tau \, \delta_0\\
&\geq K\,\left(\dfrac{s}{p}\right)^{3/2}\dfrac{\log p}{n}\geq K_* p^{-5/2},
\end{align*}
where in the third step, we used $4+\tau^2<5$ and in the final step we used $n<p$.
Substituting these bounds in \eqref{eq:split_prob}, we have for $\tau \in I_{\tau_*}$
\begin{equation}
pr(\|\beta-\beta_0\|_2<\delta \mid \tau )\geq (1-Rs/p)^{p-s}K_*e^{-(5s/2) \, \log p}
\geq e^{-Ks\log p},
\end{equation}
where $K$ is a positive constant. The proof is completed upon observing that $pr(\tau\in I_{\tau_*})\geq \tau_*/(2\pi)$, so that with a slight abuse of notation we get,
\begin{equation}
\Pi_\HS(\|\beta-\beta_0\|_2<\delta) \geq e^{-Ks\log p},
\end{equation} 
for some positive constant $K$.
\end{proof}

\vspace{-0.2in}
\section{Prior concentration results}
We establish a number of results in the following sequence of propositions and lemmas to prove Theorem 1 and 2.  The main goal here would be to establish prior concentration results around true model parameters. Recall the definitions of $\widetilde{\epsilon}_n$ and $\epsilon_n$ from Theorem 1 and 2 in the main document respectively. For Theorem 1 we need a lower bound on the prior probability assigned to the set $B_{n}^*(\eta_0,\widetilde{\epsilon}_n)=\{\eta=(C,\Sigma):\int p_{\eta_0}^{(n)}\log (p_{\eta_0}^{(n)}/p_\eta^{(n)}) dY \leq n\widetilde{\epsilon}_n^2\}$ by the product prior $\Pi_\eta =\Pi_C \otimes  \Pi_\Sigma$. Similarly, for Theorem 2, we need the prior probability of the set $B_n(C_0,\epsilon_n)=\{ C \in \Re^{p \times q}: \int p_{C_0}^{(n)} \log (p_{C_0}^{(n)}/p_{C}^{(n)})dY\leq n\epsilon_n^2,\int p_{C_0}^{(n)} \log^2 (p_{C_0}^{(n)}/p_{C}^{(n)})dY\leq n\epsilon_n^2\}$. We start by characterizing $B_{n}^*(\eta_0,\epsilon_n)$ and $B_n(C_0,\epsilon_n)$ in terms of $\| \Sigma- \Sigma_0 \|_F^2$ and $\| XC-XC_0 \|_F^2$.
\begin{proposition}\label{pr:pr1}
Consider model (3) in the main document, $Y=XC+E,\, e_i \sim \mathrm{N}(0,\Sigma)$. Then,
\begin{equation}\label{eq:KL}
\int p_{\eta_0}^{(n)} \log (p_{\eta_0}^{(n)}/p_{\eta}^{(n)})dY = \frac{n}{2}\log \frac{|\Sigma|}{ |\Sigma_0|}+\frac{n}{2}\mathrm{tr}(\Sigma^{-1}\Sigma_0 - \mathrm{I}_q) + \frac{1}{2}\| (XC-XC_0)\Sigma^{-1}(XC-XC_0)^\T \|_F^2
\end{equation}
Moreover, when $\Sigma_0=\Sigma=\mathrm{I}_q$, we have $\int p_{C_0}^{n} \log (p_{C_0}^{(n)}/p_{C}^{(n)})dY=2^{-1}\| XC - XC_0 \|_F^2$ and if $\| C -C_0 \|_F^2 < 1$, then $\int p_{C_0}^{(n)}\log^2 (p_{C_0}^{(n)}/p_C^{(n)})\leq 2^{-1} \| XC -XC_0 \|_F^2.$
\end{proposition}
\begin{proof}
The expression for $\int p_{\eta_0}^{(n)} \log (p_{\eta_0}^{(n)}/p_{\eta}^{(n)})dY$ follows directly from the formula of Kullback-Liebler divergence between two Normal distributions. Setting $\Sigma = \Sigma_0$ on the right hand side of \ref{eq:KL} yields the second assertion. Using formulas for variance and covariance of quadratic forms of Normal random vectors the third assertion is proved by noting that $\| C -C_0 \|_F^4 \leq \| C -C_0 \|_F^2$ when $\| C -C_0\|_F^2<1$.
 
\end{proof}

\begin{lemma}\label{lm:B_setup}
Let $\Sigma, \Sigma_0$ be $q\times q$ positive definite matrices and $\delta \in (0,1)$. If $\| \Sigma - \Sigma_0 \|_F \leq \delta$ and $\delta/s_{\mathrm{min}}(\Sigma_0)<1/2$, then 
$$\mathrm{tr}(\Sigma_0\Sigma^{-1} - \mathrm{I}_q) - \log \mid \Sigma_0 \Sigma^{-1} \mid \leq \frac{(K \log \rho) \delta^2}{s_{\mathrm{min}}^2(\Sigma_0)},$$
where $K$ is some absolute positive constant and $\rho = 2 s_{\mathrm{max}}(\Sigma_0)/s_{\mathrm{min}}(\Sigma_0)$. 

Furthermore,
$$\|(XC-XC_0)\Sigma^{-1}(XC-XC_0)^\T\|_F^2 \leq \{4/s_\mathrm{min}^2(\Sigma_0)\}\| XC - XC_0 \|_F^2$$. 
\end{lemma}
\begin{proof}
For the first claim see Lemma 1.3 in the supplementary document of \cite{pati2014posterior}. To prove the second claim, we use the fact the for two matrices $P$ and $Q$, $\| P Q \|_F \leq \| P \|_2 \| Q \|_F$ to get $\| (XC-XC_0)\Sigma^{-1}(XC-XC_0)^\T\|_F^2 \leq \|XC- XC_0 \|_F^2 \| \Sigma^{-1} \|_2^2$ where $\| P \|_2$ is the largest singular value of the matrix $P$. The lemma from \cite{pati2014posterior} also provides a lower bound of $s_\mathrm{min}(\Sigma)$ as $s_\mathrm{min}(\Sigma_0)/2$. Since $\| \Sigma^{-1} \|_2 = 1/s_\mathrm{min}(\Sigma_0)$, the result follows immediately. 
\end{proof} 
According to Lemma \ref{lm:B_setup}, it is equivalent to consider prior concentration of the Frobenius balls $\| XC -XC_0 \|_F$ and $\| \Sigma - \Sigma_0 \|_F$ for sufficient prior concentration around Kullback-Leibler neighborhoods. In the following sequence of Lemmas we prove $\Pi_C$ and $\Pi_\Sigma \equiv \mathrm{inv\mhyphen Wishart}(q,\mathrm{I}_q)$ satisfies such concentration.    
\begin{lemma}\label{lm:sigma_prior}
Suppose the $q \times q$ matrix $\Sigma \sim \Pi_\Sigma$ where $\Pi_\Sigma$ is the inverse-Wishart distribution with parameters $(q,\mathrm{I}_q)$. Let $\Sigma_0$ be any fixed symmetric positive definite matrix. Let $\delta$ be such that $2q^{-1/2}\delta/s_{\min}(\Sigma_0) \in (0,1)$. Then,
$$\Pi_\Sigma(\Sigma: \| \Sigma - \Sigma_0\|_F < \delta)\geq e^{-Tn\delta^2}, $$
where $T$ is a positive constant.
\end{lemma}
\begin{proof}
By the inequality $\| P Q \|_F \leq \| P \|_2 \| Q \|_F$, we have the following,
 \begin{align*}
 \Pi_\Sigma (\Sigma: \| \Sigma - \Sigma_0\|_F <\delta ) &\geq \Pi_\Sigma (\Sigma: \| \Sigma_0 \Sigma^{-1} - \mathrm{I}_q \|_F < \delta/\| \Sigma \|_2 )\\ 
 &\geq \Pi_\Sigma \{\Sigma: \| \Sigma_0 \Sigma^{-1} - \mathrm{I}_q \|_F < \delta/s_{\mathrm{min}}(\Sigma)\}\\
 &\geq \Pi_\Sigma \{\Sigma: \| \Sigma_0 \Sigma^{-1} - \mathrm{I}_q \|_F < 2\delta/ s_\mathrm{min}(\Sigma_0)\}\\
 & \geq \Pi_\Sigma \{\Sigma: \| \Sigma_0^{1/2} \Sigma^{-1} \Sigma_0^{1/2} - \mathrm{I}_q \|_F < 2\delta/ s_\mathrm{min}(\Sigma_0)\},
 \end{align*}
where we have used the the lower bound $s_{\mathrm{min}}(\Sigma) > s_{\mathrm{min}}(\Sigma_0)/2$ from the previous Lemma and the similarity of the two matrices $\Sigma_0\Sigma^{-1}$ and $\Sigma_0^{1/2}\Sigma^{-1}\Sigma_0^{1/2}$. Let $\phi_j$ be the $j^{th}$ eigenvalue of $H=\Sigma_0^{1/2}\Sigma^{-1}\Sigma_0^{1/2}$ where $j=1,\ldots,q$. Then $\| H - \mathrm{I}_q \|_F^2 = \sum_{j=1}^q (\phi_j - 1)^2 \leq 4\delta^2/s_{\mathrm{min}}^2(\Sigma_0)$. Letting $\delta_* =2 q^{-1/2} \delta/s_{\mathrm{min}}(\Sigma_0)$, we then have,
\begin{align}
&\Pi_\Sigma \bigg\{\phi_j: \sum_{j=1}^q (\phi_j -1)^2 < 4\delta^2/s_{\mathrm{min}}^2(\Sigma_0), \, j=1,\ldots, q \bigg\} \nonumber\\
&\geq \Pi_\Sigma \bigg\{\phi_j: (\phi_j -1)^2 < \delta_*^2,\, j=1,\ldots, q \bigg\} \nonumber\\
& = \Pi_\Sigma \bigg\{\phi_j: \frac{1-\delta_*}{1+\delta_*}< \phi_j < 1,\, j=1,\ldots,q \bigg\} \nonumber\\
& = \Pi_\Sigma \bigg\{\phi_j: \frac{1-\delta_*}{1+\delta_*}< \phi_j < \frac{1-\delta_*}{1+\delta_*} (1+t),\, j=1,\ldots,q \bigg\}, \label{eq:p_sigma}
\end{align} 
where $t=2\delta_*(1-\delta_*)^{-1}$ which by assumption lies in $(0,1)$. Noting that $H \sim \mathrm{Wishart}(q,\Sigma_0)$ and invoking Lemma 1 of \cite{shen2013adaptive} we have the following lower bound on the probability assigned by $\Pi_\Sigma$ to the event in \eqref{eq:p_sigma},
\begin{align}
&\Pi_\Sigma \bigg\{\phi_j: \frac{1-\delta_*}{1+\delta_*}< \phi_j < \frac{1-\delta_*}{1+\delta_*} (1+t),\, j=1,\ldots,q\bigg\}\nonumber\\
& \geq b_1 \left(\frac{1-\delta_*}{1+\delta_*}\right)^{b_2} \left(\frac{2\delta_*}{1-\delta_*}\right)^{b_3} e^{-b_4 \left(\frac{1-\delta_*}{1+\delta_*}\right)}\label{eq:p_sigma2}.
\end{align} 
Hence, for $\delta_*$ small enough, \eqref{eq:p_sigma2} can be lower bounded by $e^{-Tn\delta^2}$ for some positive constant $T$ and sufficiently large $n$.
\end{proof}

Recall the prior $\Pi_B$ from the main document. If a matrix $B \in \Re^{p \times q} \sim \Pi_B$ then each column of $B$ is a draw from $\Pi_{\HS}$. In the following Lemma we generalize Lemma 1 to provide a lower bound on the probability the prior $\Pi_B$ assigns to Frobenius neighborhoods of $B_0 \in \Re^{p \times r_0}$. By Assumption 3 the $h^{th}$ column of $B_0$, $b_h \in \ell_0[s;p]$. In order to make the Frobenius neighborhood well defined, we append $(q-r_0)$ zero columns to the right of $B_0$ and set $B_{0*}=(B_0\mid O^{p \times (q-r_0)})$.

\begin{lemma}\label{lm:B_prior}
Let the entries of $B_{0*} \in \Re^{p \times q}$ satisfy $\max \mid B_{0*} \mid \leq M$ for some positive constant $M$ . Suppose $B$ is a draw from $\Pi_B$. Define $\delta_B=\{(r_0 s \log p)/n\}^{1/2}$. Then for some positive constant $K$ we have,
$$\Pi_B(\| B-B_{0*} \|_F <\delta_B)\geq e^{-Kr_0s\log p}.$$
\end{lemma}
\begin{proof} 
Observe that, since by Assumption 3, $r_0=\kappa q$ for some $\kappa\in (0,1]$,
\begin{align*}
 \Pi_B(& \|B- B_{0*}\|_F<\delta_B)  \geq \prod_{h=1}^q\Pi_\HS\left(\|b_h-b_{0h}\|_2<q^{-1/2}\delta_B\right)\\
 &=\prod_{h=1}^q\Pi_\HS\bigg[\| b_h-b_{0h}\|_2<\{(\kappa s\log p)/n\}^{1/2}\bigg]\\
 &= \prod_{h=1}^{r_0} \Pi_\HS\bigg[\| b_h-b_{0h}\|_2<\{(\kappa s\log p)/n\}^{1/2}\bigg]\prod_{h=r_0+1}^q \Pi_\HS\bigg[\| b_h-b_{0h}\|_2<\{(\kappa s\log p)/n\}^{1/2}\bigg]\\ 
 &= \prod_{h=1}^{r_0} \Pi_\HS\bigg[\| b_h-b_{0h}\|_2<\{(\kappa s\log p)/n\}^{1/2}\bigg]\prod_{h=r_0+1}^q \Pi_\HS\bigg[\| b_h\|_2<\{(\kappa s\log p)/n\}^{1/2}\bigg].\\
 \end{align*}
From Lemma 1 we have, $\Pi_\HS[\| b_h-b_{0h}\|_2<\{(\kappa s\log p)/n\}^{1/2}]\geq e^{-K_1s\log p}$ for some positive $K_1$. Arguments along the same line of first part of Lemma 1 can also be applied to obtain that, $\Pi_\HS\left(\| b_h \|_2<\{(\kappa s\log p)/n\}^{1/2}\right)\geq (1-Rs/p)^p\geq e^{-K_2 \log p}$ for some positive $K_2$ and $R$ as defined in Lemma 1. Combining these two lower bounds in the above display we have,
$$\Pi_B(\| B-B_0\|_F<\delta_B)\geq e^{-K_1r_0s\log p} e^{-K_2(q-r_0)s\log p}.$$
Since $r_0=\kappa q$, the result follows immediately with $K=K_1+(1/\kappa-1)K_2$.
\end{proof}

Similar to the previous lemma, the following result provides a lower bound on the probability assigned to Frobenius neighborhoods of $A_0$ by the prior $\Pi_A$. Again we append $(q-r_0)$ columns at the right of $A_0$ and set $A_{0*}=(A_0 \mid O^{q \times (q-r_0)})$.
\begin{lemma}\label{lm:A_prior}
Suppose the matrix $A \sim \Pi_A$. Let $\delta_A=(qr_0/n)^{1/2}$. Then for some positive constant $K$ we have,
$$\Pi_A(\| A-A_{0*}\|_F<\delta_A)\geq e^{-Kqr_0}.$$
\end{lemma}
\begin{proof}
First we use vectorization to obtain $\Pi_A(\|A-A_0\|_F<\delta_A)=\Pi_A(\|a-a_0\|_2<\delta_A)$, where $a,a_0\in \Re^{q^2}$.
Using Anderson's lemma \citep{bhattacharya2016suboptimality} for multivariate Gaussian distributions, we then have,
\begin{align*}
\Pi_A(\| a-a_0\|_2<\delta_A)&\geq e^{-\| a_0\|^2/2}pr(\| a \|_2<\delta_A/2)\\
& = e^{-r_0/2}pr (\| a \|_2<\delta_A/2).\\
\end{align*}
The quantity $pr(\| a \|<\delta_A/2)$ can be bounded from below as,
\begin{align*}
pr(\| a \|_2 <\delta_A/2)& \geq \{pr( \mid a_j \mid <\delta_A/q)\}^{q^2}
\geq T e^{-\delta_A^2} \left(\delta_A/q\right)^{q^2}\geq e^{-Kq^2},
\end{align*}
where $K$ is a positive constant. Since $r_0=\kappa q$, it follows that $\Pi_A(\|A-A_0 \|_F<\delta_A)\geq e^{-Kqr_0}$.
\end{proof}

Our final result will concern the prior mass assigned to Frobenius neighborhoods of $C\in \Re^{p \times q}$. As in the main document we write $\Pi_C$ for prior on $C=BA^\T$ induced from $\Pi_B$ and $\Pi_A$.

\begin{lemma}\label{lm:C_prior}
Suppose $C_0$ satisfies Assumption 3. Let $C \sim \Pi_C$ with $\Pi_C$ as defined above. Define $\delta_C=\{(qr_0+r_0s\log p)/n\}^{1/2}$. Then for some positive constant $K$,
$$\Pi_C(\|C-C_0 \|_F<\delta_C)\geq e^{-K(qr_0+r_0s\log p)}.$$
\end{lemma}
\begin{proof}
Using the triangle inequality followed by the fact that for two matrices $P$ and $Q$, $\| PQ\|_F\leq s_{\max}(P)\|Q\|_F$ where $s_{\max}(P)$ is the largest singular value of $P$, we have,
\begin{align*}
\|C-C_0\|_F=\| BA^\T-B_0A_0^\T\|_F&=\| BA^\T -B_0A^\T +B_0A^\T -B_0A_0^\T\|_F\\
&=\| (B-B_{0*})A^\T+B_0(A-A_{0*})^\T\|_F\\
&\leq \|(B-B_0)A^\T\|_F+\| B_0(A-A_0)^\T\|_F\\
&\leq s_{\max}(A)\| (B-B_0)\|_F + s_{\max}(B_0) \| A-A_0\|_F,
\end{align*}
where $s_{\max}(A)$ and $s_{\max}(B_0)$ are the largest singular values of $A$ and $B_0$ respectively. From standard random matrix theory it is well known that for a random matrix of dimension $m_1 \times m_2$ with independent Gaussian entries, the largest singular admits a high probability upper bound; for every $t\geq 0$, $s_{\max(A)}\leq \sqrt{m_1}+\sqrt{m_2}+t$ with probability at least $1-2\exp{(-t^2/2)}$ \citep{vershynin2010introduction}. Also since the elements of $B_0$ are bounded, so is $s_{\max}(B_0)$, say by $\xi$. For a sufficiently large positive number $L$ and for $A\in E=\{A: s_{\max}(A)\leq 2\sqrt{q}+L\}$ we then have,

$$\|C-C_0\|_F\leq (2\sqrt{q}+L)\,\| B-B_{0*}\| _F+\xi\,\| A-A_0\|_F.$$
Thus we have, $\Pi_C(\| C-C_0\|_F<\delta_C)\geq \Pi_C\{(2\sqrt{q}+L)\| B-B_0\|_F+\xi\| A-A_0\|_F<\delta_C\}$. Since $\delta_C=\{(qr_0+r_0s\log p)/n\}^{1/2}\geq 2^{-1/2}\,[\{(r_0s(\log p)/n)\}^{1/2}+(qr_0)^{1/2}]=\\ 2^{-1/2}(\delta_B+\delta_A)$, the probability $\Pi_C(\| C-C_0\|_F<\delta_C)\geq \Pi_B(\|B-B_0\|_F< K_1 \delta_B)\,\Pi_A(\|A-A_0\|_F< K_2\delta_A)$, where $K_1$ and $K_2$ are positive constants.

From Lemma \ref{lm:B_prior} it follows that, $\Pi_B(\|B-B_0\|_F<K_1\delta_A)\geq e^{-K r_0s\log p}$ and from Lemma \ref{lm:A_prior} we have, $\Pi_A(\| A-A_0\|_F< K_2\delta_A/\beta)\geq e^{-T qr_0}$. Hence $\Pi_C(E\cap \{C:\|C-C_0\|_F<\delta_C\})\geq e^{-K(qr_0+r_0s\log p)}$. Since for two sets $E_1$ and $E_2$, $pr(E_1\cup E_2)\geq pr(E_1)+pr(E_2)-1$, the Lemma is proved.
\end{proof}

\section{Denominator in the proof of Theorem 1}

Recall $D_n$ from the proof of Theorem 1 in the main document. The following lemma establishes a high probability lower bound for $D_n$ under the true data generating distribution $P_{\eta_0}^{(n)}$. 
\begin{lemma}\label{lm:denominator}
Let $D_n= \int  e^{-\alpha r_n(\eta,\eta_0)}d\Pi_\eta$. Let $B_n^*(\eta_0,\widetilde{\epsilon_n})=\{\eta = (C,\Sigma):\int p_{\eta_0}^{(n)}( \log p_{\eta_0}^{(n)}/p_\eta^{(n)}) dY \leq n{\widetilde{\epsilon_n}}^{\,2} \}$. Then, with $P_{\eta_0}^{(n)}$-probability at least $1-K/(D-1+t)n{\widetilde{\epsilon}_n}^{\,2}$, we have,
$$D_n  \geq e^{-T\alpha (D+t)n{\widetilde{\epsilon}_n}^{\,2}},$$
for any $D>1$ and $t>0$ for some positive constants $K$.
\end{lemma}

\begin{proof}
The proof is divided into three parts. First we show that $\Pi_\eta(B_n^*\{\eta_0,\widetilde{\epsilon}_n)\}\geq e^{-Tn{\widetilde{\epsilon}_n}^{\,2}}$ for some positive $T$. Then after noting $D_n\geq \Pi_n\{B_n^*(\eta_0,\widetilde{\epsilon}_n)\}D_n^*$, where $D_n^* = \int  \int_{B_n(\eta_0,\widetilde{\epsilon_n})} e^{-\alpha r_n(\eta,\eta_0)} d\Pi^B_\eta$, we bound the expectation and variance of $Z$ where $Z$ is such that $\log D_n^* \geq Z$ and $\Pi_\eta^B$ is the restriction of $\Pi_\eta$ to $B_n^*(\eta_0,\widetilde{\epsilon}_n)$. Finally, we provide a high probability lower bound of $D_n^*$.

\vspace{0.1in}

{\bf Step 1.} From Proposition \ref{pr:pr1} if $\| \Sigma - \Sigma_0\|_F < \epsilon_n$ then $(n/2)\{\mathrm{tr}(\Sigma^{-1}\Sigma_0 - \mathrm{I}_q) - \log \mid \Sigma_0 \Sigma^{-1}\mid \}\leq n{\widetilde{\epsilon}_n}^{\,2}/2$. Also if $\| XC - XC_0 \|_F^2 \leq n\epsilon_n^2$ then $\| (XC - XC_0)\Sigma^{-1}(XC -XC_0)^\T\|_F^2 \leq n{\widetilde{\epsilon}_n}^2/2$. Hence,
$$B_n^*(\eta_0,\widetilde{\epsilon}_n) \supset A_n^*(\eta_0,\widetilde{\epsilon}_n) = \{\eta=(C,\Sigma): \| XC - XC_0 \|_F^2 \leq n{\widetilde{\epsilon}_n}^{\,2}, \, \| \Sigma - \Sigma_0\|_F < \widetilde{\epsilon}_n\},$$ 
and hence
$$\Pi_\eta \{A_n^*(\eta_0,\widetilde{\epsilon}_n)\} = \Pi_C\{C: \| XC - XC_0 \|_F^2 \leq n{\widetilde{\epsilon}_n}^{\,2} \}\,\Pi_\Sigma\{\Sigma: \| \Sigma - \Sigma_0\|_F < \widetilde{\epsilon}_n\}.$$
Using Lemma \ref{lm:sigma_prior} we get that $\Pi_\Sigma\{\Sigma: \| \Sigma - \Sigma_0 \|_F < \widetilde{\epsilon}_n\}\geq e^{-Tn{\widetilde{\epsilon}_n}^{\,2}}$. Next,
$$
\|X(C-C_0)\|_F^2\leq \|X\|_F^2 \|C-C_0\|_F^2\leq q \underset{1\leq j \leq p}{\max}\|X_j\|^2\|C-C_0\|_F^2=nq\|C-C_0\|_F^2, 
$$
where the first inequality follows from the Cauchy-Schwartz inequality and the last equality holds due to Assumption 5 for a sufficiently large $n$.
Due to Lemma \ref{lm:C_prior} we have $\Pi_C\left\lbrace C: \| C-C_0\|_F^2\leq \epsilon_n^2\right\rbrace\geq e^{-Kn\epsilon_n^2}$ for some positive constant $K$. Since $q$ is fixed, for large $n$, $nq$ is of the order $n$ and $\epsilon_n$ and $\widetilde{\epsilon}_n$ varies only by constants, therefore by the previously mentioned Lemma $\Pi_C(\| C-C_0 \|_F^2\leq\widetilde{\epsilon_n}^{2}/q)\geq e^{-Kn{\widetilde{\epsilon}_n}^{\,2}}$. Thus we get $\Pi_C\{C: \| XC - XC_0 \|_F^2 \leq n {\widetilde{\epsilon}_n}^{\,2}\}\geq e^{-Kn{\widetilde{\epsilon}_n}^{\,2}}$. Finally collecting the lower bounds for the individual probabilities we have $\Pi_\eta(B_n^*\{\eta_0,\widetilde{\epsilon}_n\})\geq e^{-Tn{\widetilde{\epsilon}_n}^{\,2}}$ for some positive $T$ as desired.

\vspace{0.1in}

{\bf Step 2.} It is obvious that, 
\begin{align*}
D_n\geq & \Pi_\eta\{B_n(\eta_0,\widetilde{\epsilon_n})\} \int_{B_n(\eta_0,\tilde{\epsilon_n})} e^{-\alpha r_n(\eta,\eta_0)} \Pi_\eta\{B_n(\eta_0,\widetilde{\epsilon}_n)\}^{-1} d\Pi_\eta\\
& = \Pi_\eta\{B_n(\eta_0,\widetilde{\epsilon}_n)\} D_{n}^*,
\end{align*}
where $D_n^* = \int_{B_n(\eta_0,\widetilde{\epsilon}_n)} e^{-\alpha r_n(\eta,\eta_0)}d\Pi_\eta^B$. Let $B$ be a shorthand for $B_n^*(\eta_0,\widetilde{\epsilon}_n)$. By Jensen's inequality applied to the concave logarithm function we then have $\log D_n^* \geq \alpha \int_B \log p_{\eta}^{(n)}/p_{\eta_0}^{(n)} d\Pi_\eta^B=Z$ (say). Then, 
\begin{align*}
  E_{\eta_0}^{(n)} (Z) = - \alpha \int_ {B}  \mathrm{KL}(p_{\eta_0}^{(n)},p_\eta^{(n)}) d\Pi_\eta^B\geq -Tn\alpha{\widetilde{\epsilon}_n}^{\,2},   
\end{align*}
for some positive $T$, where the last inequality follows from the definition of $B$.

Next we compute the variance of $Z$ under $P_{\eta_0}^{(n)}$.
\begin{align*}
\mathrm{var}_{\eta_0}(Z)& = \alpha^2 E_{\eta_0}^{(n)}\{Z - E_{\eta_0}^{(n)}(Z)\}^2\\
&= \alpha^2\int \left[\int_B \left\lbrace\log (p_{\eta_0}^{(n)}/p_\eta^{(n)}) - E_{\eta_0}^{(n)} (\log p_{\eta_0}^{(n)}/p_\eta^{(n)})\right\rbrace d\Pi_\eta\right]^2 p_{\eta_0}^{(n)} dY\\
&\leq \alpha^2 \int \int_B \left\lbrace\log (p_{\eta_0}^{(n)}/p_\eta^{(n)}) - E_{\eta_0}^{(n)} (\log p_{\eta_0}^{(n)}/p_\eta^{(n)})\right\rbrace^2 p_{\eta_0}^{(n)}d\Pi_\eta dY\\
&=\alpha^2 \int_B \left[ \int \left\lbrace\log (p_{\eta_0}^{(n)}/p_\eta^{(n)}) - E_{\eta_0}^{(n)} (\log p_{\eta_0}^{(n)}/p_\eta^{(n)})\right\rbrace^2 p_{\eta_0}^{(n)} dY\right]d\Pi_\eta\\
&=\alpha^2  \int_B \{\mathrm{var}_{\eta_0}(Z^*)\}d\Pi_\eta,
\end{align*}
where $Z^* = \log p_{\eta_0}^{(n)}/p_\eta^{(n)} = \sum_{i=1}^n \log p_{\eta_0}(Y_i)/p_{\eta}(Y_i)=\sum_{i=1}^n Z_i^*$. Hence due to independence $\mathrm{var}_{\eta_0}(Z^*)=n\mathrm{var}_{\eta_0}(Z_1^*)$. Now,
\begin{align*}
\mathrm{var}_{\eta_0}(Z_1^*) = \mathrm{var}_{\eta_0}\left\lbrace\frac{1}{2}(Y_1 - C^\T x_1)\Sigma^{-1}(Y_1-C^\T x_1)-\frac{1}{2}(Y_1 - C_0^\T x_1)\Sigma_0^{-1}(Y_1-C_0^\T x_1)\right\rbrace.
\end{align*}
Let $(Y_1-C_0^\T x_1)=u_0$ and $(Y_1-C^\T x_1)= u_1$ and $(C_0^\T x_1 - C^\T x_1)=u$. Then,
 \begin{align*}
 \mathrm{var}_{\eta_0}(Z_1^*)& =\frac{1}{4}\mathrm{var}_{\eta_0}\left(u_1^\T \Sigma^{-1}u_1 - u_0^\T \Sigma^{-1}u_0 + u_0^\T \Sigma^{-1}u_0 - u_0^\T \Sigma_0^{-1}u_0\right)\\
 &=\frac{1}{4}\mathrm{var}_{\eta_0}\left\lbrace u^\T \Sigma^{-1} u + 2 u^\T \Sigma^{-1} u_0 + u_0^\T (\Sigma^{-1} - \Sigma_0^{-1}) u_0 \right\rbrace\\
 & = \frac{1}{4}\mathrm{var}_{\eta_0} \left\lbrace 2 u^\T \Sigma^{-1} u_0 + u_0^\T (\Sigma^{-1} - \Sigma_0^{-1}) u_0\right\rbrace \\
 & = \mathrm{var}_{\eta_0}(u^\T \Sigma ^{-1} u_0) + \frac{1}{4} \mathrm{var}_{\eta_0}\left\lbrace u_0^\T (\Sigma^{-1} - \Sigma_0^{-1})u_0\right\rbrace +\frac{1}{2} \mathrm{cov}_{\eta_0}\left\lbrace u^\T \Sigma^{-1}u_0  , u_0^\T (\Sigma^{-1} - \Sigma_0^{-1})u_0\right\rbrace\\
& = u^\T \Sigma^{-1} \Sigma_0 \Sigma^{-1} u + \frac{1}{2} \mathrm{tr}\{(\Sigma^{-1}\Sigma_0 - \mathrm{I}_q)^2\} \\
& \leq \| C_0^\T x_1 - C^\T x_1 \|_2^2\,  \| \Sigma_0 \|_2 \| \Sigma^{-1} \|_2^2 + \frac{1}{2} \| \Sigma^{-1/2} \Sigma_0 \Sigma^{-1/2} - \mathrm{I}_q \|_F^2\\
& \leq \| C_0^\T x_1 - C^\T x_1 \|_2^2\,  \| \Sigma_0 \|_2 \| \Sigma^{-1} \|_2^2 + \frac{1}{2} \| \Sigma_0  - \Sigma \|_F^2 \| \Sigma ^{-1} \|_2^2\\
& \leq \frac{4 \| \Sigma_0 \|_2}{s_{\min}^2(\Sigma_0)}\| C_0^\T x_1 - C^\T x_1 \|_2^2\,   + \frac{1}{2} \| \Sigma^{-1} \|_2^2  \| \Sigma - \Sigma_0 \|_F^2\\
&\leq  \frac{4 \| \Sigma_0 \|_2}{s_{\min}^2(\Sigma_0)}\| C_0^\T x_1 - C^\T x_1 \|_2^2\,   + \frac{2}{s_{\min}^2(\Sigma_0)}  \| \Sigma - \Sigma_0 \|_F^2,
 \end{align*}
 where we have used the lower bound on $s_{\min}(\Sigma)$ from Lemma \ref{lm:sigma_prior}.
Therefore, $\mathrm{var}_{\eta_0}(Z^*)\leq \alpha^2\frac{4 \| \Sigma_0 \|_2}{s_{\min}^2(\Sigma_0)} \| XC - XC_0 \|_F^2 + \alpha^2 \frac{2n}{s_{\min}^2(\Sigma_0)} \| \Sigma - \Sigma_0 \|_F^2$. Since $B \supset A_n^*(\eta_0,\widetilde{\epsilon}_n)$ from step 1, and $\Pi\{A_n^*(\eta_0,\widetilde{\epsilon}_n)\}$, we finally get $\mathrm{var}_{\eta_0}(Z)\leq K\alpha^2 n{\widetilde{\epsilon}_n}^{
\,2}$ for some positive constant $K$.

\vspace{0.1in}

{\bf Step 3.} For any $D>1$ and $t>0$, by Chebyshev's inequality
\begin{align*}
P_{\eta_0}^{(n)}\{Z \leq -T \alpha(D+t)n{\widetilde{\epsilon}_n}^{\,2}\}& = P_{\eta_0}^{(n)}\{Z \leq -T \alpha(D-1+t+1)n{\widetilde{\epsilon}_n}^{\,2}\}\\
& = P_{\eta_0}^{(n)}\{Z - (-T\alpha n{\widetilde{\epsilon}_n}^{\,2})\leq -T \alpha(D-1+t)n\widetilde{\epsilon}_n^2\}\\
&\leq \frac{\mathrm{var}_{\eta_0}(Z)}{\left\lbrace T \alpha(D-1+t)n{\widetilde{\epsilon}_n}^{\,2}\right\rbrace^2}\\
& \leq \frac{K}{T^2(D-1+t)^2 n{\widetilde{\epsilon}_n}^{\,2}},
\end{align*}
where we have used the fact that $\mathrm{var}_{\eta_0}(Z)\leq Kn{\widetilde{\epsilon}_n}^{\,2}$. Thus we get with $P_{\eta_0}^{(n)}$-probability at least $1-K/(D-1+t)^2 n{\widetilde{\epsilon}_n}^{\,2}$,
$$\log D_n^* \geq -T\alpha(D+t)n{\widetilde{\epsilon}_n}^{\,2} \Leftrightarrow D_n^* \geq e^{-T\alpha(D+t)n{\widetilde{\epsilon}_n}^{\,2}},$$
for some positive constant $K$. Since $D_n \geq \Pi_\eta(B)D_n^*$ and $\Pi_{\eta}(B)\geq e^{-Tn\widetilde{\epsilon}_n^2}\geq e^{-T\alpha(D+t)n\widetilde{\epsilon}_n^2} \, (D>1)$, we finally obtain,
$$D_n\geq e^{-T\alpha(D+t)n\widetilde{\epsilon}_n^2},$$
with $P_{\eta_0}^{(n)}$-probability at least $1-K/(D-1+t)^2 n\widetilde{\epsilon}_n^2$ for some positive constant $K$. 
\end{proof}

\section*{Proof of Theorem 3}
\begin{proof}
 
\noindent Let $m_\alpha(Y) = \int \{p^{(n)}(Y \mid C, \Sigma ; \, X)\}^\alpha \, \Pi_C(dC) \Pi_\Sigma(d\Sigma)$ where $\alpha \in (0,1)$ and $m(Y) = \int p^{(n)}(Y \mid C, \Sigma; \, X) \, \Pi_C(dC) \Pi_\Sigma (d\Sigma)$. Also, $m(Y) >0$ for every $Y$ due to the positivity of $p^{(n)}(Y \mid C, \Sigma; \, X)$ and $\Pi_C(dC)\Pi_\Sigma(d\Sigma)$. Furthermore, for every $(C,\Sigma)$, $\lim_{\alpha \to 1_-} \{p^{(n)}(Y \mid C,\Sigma ; \, X)\}^\alpha \, \Pi_C(dC)\Pi_\Sigma(d\Sigma) = p^{(n)}(Y \mid C,\Sigma ; \, X) \, \Pi_C(dC)\Pi_\Sigma(d\Sigma)$. Since $\{p^{(n)}(Y \mid C ; \, X)\}^\alpha \, \Pi_C(C) \leq \Pi_{dC}\Pi_{\Sigma}(d\Sigma)$ and $\int \Pi_C(dC)\Pi_{\Sigma}(d\Sigma) = 1$, by the dominated convergence theorem $\lim_{\alpha \to 1_{-}} m_{\alpha}(Y) = m(Y)$. Combining, we get that $\lim_{\alpha \to 1_-}\Pi_{n,\alpha}(C,\Sigma \mid Y) = \Pi_n(C,\Sigma \mid Y)$ for all $(C,\Sigma)$.
Then by Scheffe's theorem we get the desired result.
\end{proof}

\section*{Proof of Theorem 4}
We first prove the following Lemma related to prior concentration of an $\mbox{inverse-Gamma}(n(1-\alpha)/2 +a, \alpha b)$ prior where $\alpha = \{1- 1/(\log n)^t\}, t>1$.
\begin{lemma}\label{lm:new_sigma_prior}
Let $\tau^2 \sim IG(n(1-\alpha)/2 + a, \alpha b)$ for some fixed $a , b >0$ and $\alpha = \{1- 1/(\log n)^t\}, t>1$.  Then for any fixed $\sigma_0^2 > 0$ and $\epsilon >0$
$$P[\,\,\vert \tau^2 - \sigma_0^2 \vert < \epsilon] \geq e^{-Cn \epsilon},$$
for some positive $C$.
\end{lemma}
\begin{proof}
Without loss of generality let $\sigma_0^2 = 1$. Since otherwise $P[\,\,\vert \tau^2 - \sigma_0^2 \vert < \epsilon] = P[\,\,\vert \tau^2/\sigma_0^2 - 1 \vert < \epsilon/\sigma_0^2] = P[\,\,\vert \tau_*^2 - 1 \vert < \delta],$ where $\tau_*^2 \sim IG(n(1-\alpha)/2+ a, \alpha b/\sigma_0^2 )$ and $\delta = \epsilon/\sigma_0^2$ is fixed.

We have, 
\begin{align*}
\Pi(\,\,\vert \tau^2 - 1\vert\,\,<\epsilon) &\geq \Pi(1 < \tau^2 < 1+\epsilon)\\
& = \dfrac{(b\alpha)^{n(1-\alpha)/2 + a}}{\Gamma \{n(1-\alpha)/2 + a\}} \int_{1}^{1+\epsilon} (\tau^2)^{- n(1-\alpha)/2 - a - 1} \exp{(-b\alpha/\tau^2)} d\tau^2 \\
& \geq \dfrac{(b\alpha)^{n(1-\alpha)/2 + a}}{\Gamma \{n(1-\alpha)/2 + a\}} \exp{(-b\alpha)} \int_{1}^{1+\epsilon} (\tau^2)^{- n(1-\alpha)/2 - a - 1}d\tau^2\\
& \geq \dfrac{e^{-b}b^a e^{n(1-\alpha)/2 \log b} e^{a \log \alpha}e^{n(1-\alpha)/2 \log \alpha }}{\Gamma n(1-\alpha)} \int_{1}^{1+\epsilon} (\tau^2)^{- n(1-\alpha)/2 - a - 1}d\tau^2\\
&\geq \dfrac{C e^{n(1-\alpha)/2 \log b} e^{a \log \alpha}e^{n(1-\alpha)/2 \log \alpha}}{\Gamma n(1-\alpha)} \dfrac{[1 - (1+\epsilon)^{n(1-\alpha)/2 + a}]}{n(1-\alpha)/2 + a}\\
&\geq \dfrac{C e^{n(1-\alpha)/2 \log b} e^{a \log \alpha}e^{n(1-\alpha)/2 \log \alpha}}{\Gamma n(1-\alpha)} \epsilon,
\end{align*}
where in the last step we have used $(1+x)^n \leq nx, x\in (0,1)$. Using Stirling's approximation we get $\Gamma n(1-\alpha) = \{2\pi/n(1-\alpha))^{1/2}\} \{n(1-\alpha)/e\}^{n(1-\alpha)}$. Putting this together in the above expression we get the following lower bound,
$$\Pi(\,\,\vert \tau^2 - 1\vert\,\,<\epsilon) \geq e^{-C n(1-\alpha)\log \{n(1-\alpha)\}},$$
for some positive $C$. Now for $\alpha = \{1 - 1/(\log n)^t\}$ we have $n(1-\alpha)\log \{n(1-\alpha)\} = n/(\log n)^{t - 1} \leq n\epsilon$ for large $n$ and fixed $\epsilon > 0$.
\end{proof}

We are now ready to prove Theorem 4. Recall from the main document that $\Sigma_* = \alpha \Sigma = \alpha \mbox{diag}(\sigma_1^2, \ldots, \sigma_q^2)$ and $\Pi_{\Sigma_*} = \prod_{h=1}^q \Pi_{\tau_h^2}$ where  $\tau_h^2 = \alpha \sigma_h^2$ and $\Pi_{\tau_h^2} \equiv \mbox{inverse-Gamma}\{n(1-\alpha)/2 +a, b\alpha\}$

\begin{proof}
For any $\alpha \in (0,1)$, as noted in the main document, 
\begin{align*}
\Pi_{n}(C, \Sigma \mid Y)& \propto |\Sigma|^{-n/2} \, e^{- \mbox{tr}\{ (Y - X C)\Sigma^{-1} (Y - X C)^{\T} \}/2} \Pi_C(dC) \Pi_\Sigma(d\Sigma)\\
& \propto |\Sigma_*|^{-n\alpha/2} \, e^{- \alpha \mbox{tr}\{ (Y - X C)\Sigma_*^{-1} (Y - X C)^{\T} \}/2} \Pi_C(dC) \Pi_{\Sigma_*}(d\Sigma_*)\\
& \propto \Pi_{n,\alpha}(C, \Sigma_* \mid Y)
\end{align*} 
where $\Sigma_* = \alpha \Sigma$ and $\Pi_{\Sigma_*}(\cdot)$ is again a product with components $\mbox{inverse-Gamma}\{n(1-\alpha)/2 + a, \alpha b\}$. Since the first and last terms in the above displays are both probability densities, we conclude that $\Pi_{n}(C, \Sigma \mid Y) =  \Pi_{n,\alpha}(C, \Sigma_* \mid Y)$. 

Set $\alpha = 1-1/(\log n)^t$ for $t > 1$ large enough. With this choice, we shall show consistency of $\Pi_{n,\alpha}(C, \Sigma_* \mid Y)$, which in turn will imply consistency of $\Pi_{n}(C, \Sigma \mid Y)$ in the average Hellinger metric.

 
The fractional posterior probability of a set $B_n$ is given by,
\begin{equation}
T_n = \Pi_{n,\alpha}(B_n \mid Y) = \dfrac{\int_{B_n}e^{-\alpha r_n(P,P_0)}\Pi_C(dC)\Pi_{\Sigma_*}(d\Sigma_*)}{\int e^{-\alpha r_n(P,P_0)}d\Pi_C(dC)\Pi_{\Sigma_*}(d\Sigma_*)},
\end{equation}
where $r_n(P,P_0) = \sum_{i=1}^n \{\log p_0(y_i)/p(y_i)\}$ with $p$ and $p_0$ being the respective densities of $P$ and $P_0$. Under $P_0$, $y_i \sim \Gauss(C^\T x_i, \Sigma)$ and under $P$, $y_i \sim \Gauss(C_0^\T x_i, \Sigma_0)$. Call the numerator in the above display $N_n$ and the denominator $D_n$. 

For $\Pi_{n,\alpha}(\cdot\mid Y)$ to be consistent we need condition 3 of Theorem \ref{th:ghosal} to hold for any given $\epsilon > 0$. Also due to Lemma \ref{lm:B_setup} this reduces to showing prior concentration for balls of type $\norm{XC - XC_0}_F^2$ and $\norm{\Sigma_* - \Sigma_{0}}_F^2 = \sum_{h = 1}^q (\tau_h^2 - \sigma_{0h}^2)^2$. For any fixed $\epsilon>0$, we already have $\Pi_C\{C: \norm{XC - XC_0}_F^2 < \epsilon\}\geq e^{-K_1n\epsilon}$ from Step 1 of Lemma \ref{lm:denominator} for some constant $K_1>0$. 
Furthermore, $\Pi_{\Sigma_*}\{\sum_{h=1}^q (\tau_h^2 - \sigma_{0h}^2)^2 < \epsilon\}\geq \Pi_{\Sigma_*}\{|\tau_h^2 - \sigma_{0h}^2|<\epsilon/q, h = 1,\ldots, q\} = \prod_{h = 1}^q \Pi (\, \, \vert \tau_h^2 - \sigma_{0h}^2 \vert \,\, < \epsilon)\geq e^{-K_2 n\epsilon}$ due to Lemma \ref{lm:new_sigma_prior}. Thus if $B = \{p: \int p_0\log (p_0/p) < \epsilon\}$, then $\Pi(B) = (\Pi_C \otimes \Pi_{\Sigma_*})(B)>e^{-Kn \epsilon}$ for some positive $K$.

For $D_n$ we follow standard arguments \citep{ghosal2017fundamentals} to provide it with the following lower bound adapted for fractional posteriors
$$D_n \geq \Pi(B)e^{-n\alpha \epsilon} \geq e ^{-K_0 n \epsilon}, \quad \mbox{for some } K_0 >0,$$
where $\alpha = 1 - 1/(\log n)^t$ and $n\alpha < n$.

Now set $B_n =\{D_\alpha(p,p_0)>Mn\epsilon\} = \{D_\alpha(p,p_0)> M n \epsilon\}$ for large $M>0$ where $D_\alpha(p, p_0)$ is the R\'enyi divergence of order $\alpha$.
Then $\mathbb{E}_{P_0}(N_n) < e^{-M n \epsilon/(\log n)^t} \leq e^{-Mn\epsilon}$ following arguments from \cite{bhattacharya2016bayesian}. Thus $\mathbb{E}_{P_0}(T_n)\leq e^{-Mn \epsilon}/e^{-K_0 n \epsilon}\}\leq e^{-M_0 n\epsilon}$ for suitably large $M$ and $M_0>0$. Now for any $\delta >0$, by Markov's inequality
\begin{equation}
\sum_{n}P(T_n > \delta)\leq \delta^{-1}\sum_n e^{-M_0n \epsilon}<\infty,
\end{equation} 
Hence $\Pi_{n,\alpha}(\cdot \mid Y)$ is consistent by the Borel--Cantelli lemma and thus $\Pi_n(\cdot \mid Y)$ is also consistent. Using the equivalence between R\'enyi divergences and the Hellinger distance between densities the statement of the Theorem is now proved.
\end{proof}

\section*{Derivation of equations from section 3.1 in the main document}
\subsection*{Derivation of equation (8)}
Set $\Sigma=\mathrm{I}_q$. Suppose $Y^* \in \Re^{n \times q}$ be $n$ future observations with design points $X$ so that given $C$, $Y^*$ can be decomposed into $Y^*=XC+E^*$ where $E^*$ where the individual rows of $E^*$ follow $\mathrm{N}(0,\Sigma)$. We define the utility function in terms of loss of predicting these $n$ new future observations. To encourage sparsity in rows of a coefficient matrix $\Gamma$ that balances the prediction we add a group lasso penalty \citep{yuan2006model} to this utility function. We define the utility function as,

\begin{equation}\label{eq:utility10}
\mathcal{L}(Y^*,\Gamma)=\| Y^*-X\Gamma \|_F^2 + \sum_{j=1}^p \mu_j \| \Gamma^{(j)} \|_2
\end{equation}
where the $p$ tuning parameters $\{\mu_j\}_{j=1}^p$ control the penalty for selecting each predictor variable and $\Phi^{(j)}$ represents the $j^{th}$ row of any matrix $\Phi$. Intuitively we want $\mu_j$ to be small if the $j^{th}$ predictor is important and vice versa.
The expected risk, $\mathbb{E}\{\mathcal{L}(Y^*,\Gamma)\}$, after integrating over the space of all such future observations given $C$ and $\Sigma$, is 
\begin{align}
\mathcal{L}(\Gamma,C,\Sigma)=q \text{ tr}(\Sigma)+ \| XC-X\Gamma \|_F^2 + \sum_{j=1}^p \mu_j \| \Gamma^{(j)} \|_2.
\end{align}

Finally we take expectation of this quantity with respect to $\pi(C \mid Y,X)$ and drop the constant terms to obtain (9).

\subsection*{Derivation of equation (9)}
We let $\Phi_j$ and $\Phi^{(j)}$ denote the $j^{th}$ column and row of a generic matrix $\Phi$. Using the subgradient of (10) with respect to $\Gamma^{(j)}$ \citep{friedman2007pathwise}, we have 
\begin{equation}\label{eq:subgrad}
2X_j^\T(X\Gamma-X\overline{C})+\mu_j \alpha_j=0, \quad j=1,\ldots,p,
\end{equation}
where $\alpha_j=\Gamma^{(j)}/\| \Gamma^{(j)} \|$ if $\| \Gamma^{(j)} \| \neq 0$ and $\| \alpha_j \| < 1$ when $\| \Gamma^{(j)} \|=0$. For $\Gamma^{(j)}=0$ we can rewrite \eqref{eq:subgrad} as, $2X_j^\T(\sum_{k\neq j}X_k\Gamma^{(k)}-X\overline{C})+\mu_j \alpha_j=0$ which imply that
$\alpha_j=-2X_j^\T R_j/\mu_j$, where $R_j$ is the residual matrix obtained after regressing $X\overline{C}$ on $X$ leaving out the $j^{th}$ predictor, $R_j=X\overline{C}-\sum_{k\neq j}X_k \Gamma^{(k)}$. We can use this to set $\Gamma^{(j)}$ to zero: if $\alpha_j<1$ set $\Gamma^{(j)}=0$. Otherwise we have $2X_j^\T(X_j\Gamma^{(j)}-R_j)+\mu_j \Gamma^{j}/\|\Gamma^{(j)} \|=0$. Solving for $\Gamma^{(j)}$ in the above equation we then get,
\begin{equation}\label{eq:subgrad1}
\Gamma^{(j)}=\left(X_j^\T X_j+\dfrac{\mu_j}{2\| \Gamma^{(j)} \|}\right )^{-1} X_j^\T R_j.
\end{equation} 
This solution is dependent on the unknown quantity $\| \Gamma^{(j)} \|$. However, taking norm on both sides in \eqref{eq:subgrad1} we get a value of $\|\Gamma^{(j)} \|$ which does not involve any unknown quantities: $\| \Gamma^{(j)} \|=(\| X_j^\T R_j \|-\mu_j/2)/X_j^\T X_j$. Substituting this in \eqref{eq:subgrad1} we get, $\Gamma^{(j)}=(X_j^\T X_j)^{-1}\left(1-\mu_j/2\| X_j^\T R_j \|\right ) X_j^\T R_j$.  

Finally, combining the case when $\Gamma^{(j)}=0$, we have (10).

\section*{Yeast cell cycle data}
The yeast cell cycle data consists of mRNA measurements $Y$, measured every 7 minutes in a period of 119 minutes. The covariates $X$ are binding information on 106 transcription factors. When applied to this data, the proposed method identified 33 transcription factors out of 106 that driving the variation in mRNA measurements. 14 of the identified transcription factors are among the 21 scientifically verified \citep{lee2002transcriptional}. In the main document we provided estimated effects of two of the 21 scientifically verified transcription factors. Here we plot the estimated effects of the remaining transcriptions factors that were scientifically verified.
\begin{figure}
\begin{flushleft}
\caption{Estimated effects of the 19 of 21 scientifically verified transcription factors selected by the proposed method. Effects of other two, viz. ACE2 and SWI4 are included in the main manuscript. Red lines correspond to 95\% posterior symmetric credible intervals, black lines represent the posterior mean and the blue dashed line plots values of the BSML estimate $\widehat{C}_{RR}$.}
\label{fig:yeast2}
\includegraphics[height=6in,width=6in]{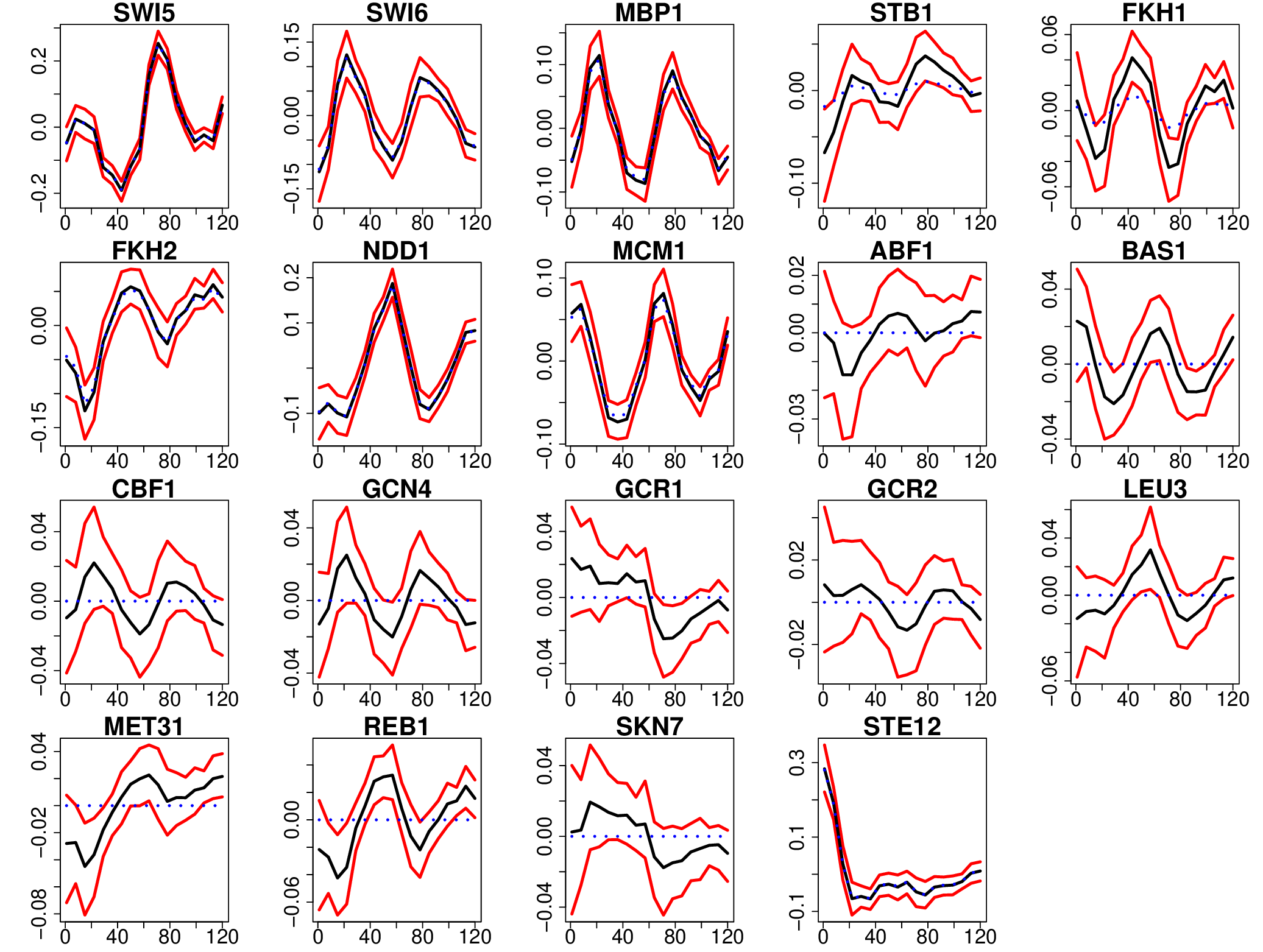}
\end{flushleft}
\end{figure}
\end{document}